\documentclass[lettersize,journal]{IEEEtran}
\usepackage{fontawesome5}
\usepackage{amsmath,amsfonts,amssymb}
\usepackage{array}
\usepackage[caption=false,font=normalsize,labelfont=sf,textfont=sf]{subfig}
\usepackage{textcomp}
\usepackage{stfloats}
\usepackage{url}
\usepackage{graphicx}
\usepackage{xcolor}
\usepackage{cite}
\usepackage{booktabs}
\usepackage{multirow}
\usepackage{bm}
\usepackage{tikz}
\usepackage{pgfplots}
\usepackage{standalone}
\usepackage{amsmath}
\usepackage{xcolor}
\usepackage{comment}

\usepackage{booktabs}

\usepackage{pgfplots}
\pgfplotsset{compat=1.18}
\usepgfplotslibrary{fillbetween} 
\usepgfplotslibrary{groupplots}
\usepgfplotslibrary{statistics}

\definecolor{rwthblue100}{HTML}{00549F}
\definecolor{rwthblue50}{HTML}{8EBAE5}
\definecolor{rwthpetrol100}{HTML}{0F4D58}
\definecolor{rwthgray75}{HTML}{7F7F7F}
\definecolor{rwthred100}{HTML}{CC071E}

\definecolor{bglightgreen}{HTML}{E8F5E9}
\definecolor{bglightpurple}{HTML}{E8EAF6}
\definecolor{bglightblue}{HTML}{E3F2FD}
\usepackage{pgfplots}
\usepackage{pgfplotstable}
\usetikzlibrary{calc}
\usepgfplotslibrary{groupplots}
\pgfplotsset{compat=1.18}

\definecolor{rwthblue100}{HTML}{00549F}
\definecolor{rwthblue50}{HTML}{8EBAE5}
\definecolor{rwthpetrol100}{HTML}{0F4D58}
\definecolor{rwthgray75}{HTML}{7F7F7F}
\definecolor{rwthred100}{HTML}{CC071E}

\usepackage{booktabs}
\usepackage{hyperref}
\usepackage[table]{xcolor}
\definecolor{sGreen}{RGB}{0,150,0}
\definecolor{sRed}{RGB}{200,0,0}

\usepackage{amsmath}
\usepackage{amssymb}

\usetikzlibrary{positioning, shapes.geometric, arrows.meta, calc, backgrounds, matrix, patterns, fit}

\definecolor{colorBgOffline}{RGB}{245, 245, 245}  
\definecolor{colorBgOnline}{RGB}{235, 245, 252}   
\definecolor{colorArrow}{RGB}{86, 180, 233}       
\definecolor{colorSafe}{RGB}{0, 158, 115}         
\definecolor{colorSafeBg}{RGB}{225, 247, 238}     
\definecolor{colorWarn}{RGB}{213, 94, 0}          
\definecolor{colorWarnBg}{RGB}{254, 238, 228}     
\definecolor{colorPlotPink}{RGB}{230, 159, 0}     
\definecolor{colorPlotGreen}{RGB}{0, 114, 178}    

\definecolor{cblue}{HTML}{2563EB}
\definecolor{cteal}{HTML}{0F766E}
\definecolor{corange}{HTML}{B45309}
\definecolor{cgray}{HTML}{475569}
\definecolor{cdark}{HTML}{0F172A}
\definecolor{cred}{HTML}{DC2626}
\definecolor{cbar}{HTML}{93C5FD}
\definecolor{cblue}{HTML}{2563EB}
\definecolor{cteal}{HTML}{0F766E}
\definecolor{cpurple}{HTML}{7C3AED}
\definecolor{cgreen}{HTML}{15803D}
\definecolor{corange}{HTML}{B45309}
\definecolor{cgray}{HTML}{475569}
\definecolor{cdark}{HTML}{0F172A}
\colorlet{unmodeled_color}{orange!85!black}
\colorlet{conformalblue}{blue!18}
\colorlet{vividorange}{orange!90!black}
\usetikzlibrary{arrows.meta,positioning,calc,fit,backgrounds,shadows}
\pgfplotsset{compat=1.18}

\definecolor{unmodeled_color}{RGB}{120, 0, 200}
\definecolor{vividorange}{RGB}{255,100,0}
\definecolor{conformalblue}{RGB}{200,225,255}

\definecolor{rwthblue100}{HTML}{00549F}
\definecolor{rwthblue50}{HTML}{8EBAE5}
\definecolor{rwthpetrol100}{HTML}{0F4D58}
\definecolor{rwthgray75}{HTML}{7F7F7F}
\definecolor{rwthred100}{HTML}{CC071E}
\definecolor{bglightgreen}{HTML}{DDEFE8}
\definecolor{bglightpurple}{HTML}{E9E3F5}
\definecolor{bglightblue}{HTML}{DDEBF7}

\newcommand{\safeincludegraphics}[2][]{%
  \IfFileExists{#2}{%
    \includegraphics[#1]{#2}%
  }{%
    \fbox{%
      \begin{minipage}[c][0.18\textheight][c]{0.92\linewidth}
      \centering
      Figure placeholder\\[1mm]
      \texttt{#2}
      \end{minipage}%
    }%
  }%
}
\renewcommand{\baselinestretch}{0.985}

\begin{document}

\title{Safety Screening for Voltage Control in Active Distribution Grids \\
via Distributionally Robust Conformal Screening}

\author{Sarra~Bouchkati\textsuperscript{*,\textdagger},
        Petros~Ellinas\textsuperscript{*,\textdagger},
        Adriana~Geisler\textsuperscript{*,\textdagger},
        Steffen~Kortmann,
        Johanna~Vorwerk,
        Spyros~Chatzivasiliadis,
        and~Andreas~Ulbig%
\thanks{\textsuperscript{*}contributed equally to this work.}
\thanks{\textsuperscript{\textdagger}Corresponding authors: Sarra Bouchkati (e-mail: s.bouchkati@iaew.rwth-aachen.de), Petros Ellinas (e-mail: petrel@dtu.dk), and Adriana Geisler (e-mail: a.geisler@iaew.rwth-aachen.de).}
\thanks{Sarra Bouchkati, Adriana Geisler, Steffen Kortmann, and Andreas Ulbig are with IAEW at RWTH Aachen University, Aachen, Germany.}
\thanks{Petros Ellinas, Johanna Vorwerk, and Spyros Chatzivasiliadis are with the Department of Wind and Energy Systems, Technical University of Denmark, Kongens Lyngby, Denmark.}
\thanks{The work of P. Ellinas and S. Chatzivasileiadis is supported by the European Research Council (ERC) Starting Grant VeriPhIED, Grant Agreement No.~949899.}
}

\markboth{IEEE Transactions on Smart Grid,~Vol.~xx, No.~xx, Month~20xx}%
{Author \MakeLowercase{\textit{et al.}}: DR-CSS for Learning-Based Voltage Control}

\maketitle

\begin{abstract}
Deploying a new control policy for voltage control in active distribution grids requires evidence that physical limits will be satisfied before the policy is tested on the physical grid. This assessment is difficult for two reasons. First, simulations cannot capture every disturbance, modeling error, and device interaction present in the real grid. Second, historical measurements reflect operation under existing control policies, whereas a new policy may drive the grid into different operating conditions.
To address these challenges, we propose \emph{Distributionally Robust Conformal Safety Screening (DR-CSS)}, a policy-agnostic framework for pre-deployment, scenario-by-scenario screening of a new control policy using historical data and a nominal simulator. For each new scenario, the simulator predicts a future voltage trajectory for the whole grid; DR-CSS then constructs a conformal safety interval around this prediction using historical simulation-to-reality errors. The interval is further enlarged to account for closed-loop changes induced by the deployment of the new policy and its interactions with the remaining controllers. To the best of our knowledge, DR-CSS is the first framework in power systems to combine historical data from an existing control policy with an imperfect simulator for pre-deployment safety screening of a new policy. 
Experiments on the IEEE 33-bus and IEEE 141-bus systems evaluate the deployment of learning-based voltage control policies and show that DR-CSS identifies all unsafe test scenarios. To reduce unnecessary warnings on safe scenarios, we adapt the safety intervals to different operating conditions and gradually introduce new policies with recalibration after each stage. These extensions increase the informational value of the safety screening and support safer deployment decisions in active distribution grids.

\end{abstract}


\section{Introduction}

One of the growing challenges in operating active distribution grids is maintaining bus voltages within permissible limits. Numerous inverter-interfaced distributed energy resources (DERs), flexible loads, and heterogeneous control devices operate as independent decision-making components, forming a complex \textit{multi-agent} environment that shapes real-time grid operation. While the intermittency of renewable DERs increases the risk of voltage-limit violations \cite{MURRAY2021111100}, smart inverters can help regulate voltages through their fast and continuously adjustable reactive-power capabilities \cite{CUI2022108609}. Coordinating reactive-power injections to maintain acceptable voltage profiles, often while limiting control effort or associated costs, is commonly referred to as the Volt-Var control problem. Various control policies, or automated decision-making rules, have been proposed to address this problem, ranging from simple rule-based local droop schemes to advanced learning-based methods such as reinforcement learning (RL)\cite{ZHA21,Wang2020SafeOffPolicy,Gao2021ConsensusMARL,Kabir2023TwoTimescaleVVC,Wang2021}. However, because these policies often act solely based on only partial grid information, they may fail to guarantee grid-wide safe voltage control due to the lack of coordination \cite{Saverio2019}. Several approaches aimed at addressing safety at the control policy-design stage. Safe RL techniques embed safety layers or constraint-satisfaction mechanisms directly into the training process to minimize operational risk, typically at the cost of more conservative exploration, or degraded nominal performance \cite{Wang2020SafeOffPolicy,Gao2022ModelAugmentedSafeRL,Nguyen2022OnlineSafeDRL}. Crucially, these methods aim to enforce safety during policy training, but their guarantees are typically tied to the assumed training environment. They therefore do not guarantee grid-wide safety after deployment, where interactions with the physical grid and other controllers may lead to unanticipated voltage violations.

Deploying a new control policy at one or more inverters changes its interactions with the grid and
with the controllers that remain in operation. Therefore, a key open problem is \emph{how to assess, before deployment,
whether a new control policy introduced at one or more inverter buses will
maintain all bus voltages within their admissible limits, using historical
data and an imperfect simulator and without testing unsafe conditions on
the physical system.}

Ensuring the safe deployment of new control policies involves two complementary
procedures: validation, which evaluates performance through simulation and
empirical testing, and analytical verification, which provides analytical safety
guarantees. Both are challenging in distribution grids because grid operators
often lack a sufficiently accurate model of the physical system. In
practice, distribution-grid models are limited by incomplete observability,
data-quality issues, and imperfect topology knowledge
\cite{Geth2023DSSEPractical,Parlaktuna2024TopologyError}, as well as
uncertainties in renewable injections, load behaviour, and measurement errors \cite{Wang2025UncertaintyReview}. Traditional
validation tools, such as scenario simulation and hardware-in-the-loop
testing \cite{Wang2019PHIL,OTT18,PAS24}, therefore remain conditional on model fidelity and scenario representativeness. Direct testing of potentially unsafe conditions on the physical grid is also unacceptable, since power grids constitute a safety-critical infrastructure. Furthermore, historical operational data provides only limited support for evaluating new control policies, since it is generated under the closed-loop behavior of previously deployed policies. Consequently, it may not adequately represent the states, interactions, and control actions that would arise after an untested control policy is introduced.

Analytical verification workflows share a common modeling bottleneck: their guarantees are only as reliable as the models on which they are built. Formal verification, reachability analysis, and certificate-based methods
can provide strong deterministic guarantees under suitable assumptions, but
they typically require accurate system models, tractable mathematical
abstractions, and substantial computational resources
\cite{Tran2022VerificationLECPS,Everett2021NNVerification,
Kochdumper2023OpenClosedLoop,Dawson2023LearnedCertificates}. When the underlying model deviates from the true system, these guarantees no longer transfer to physical deployment.

Statistical uncertainty quantification offers a complementary way to assess
new control policies before deployment. Existing approaches include probabilistic voltage-risk
assessment and conformal prediction for power-system forecasts and voltage
trajectories
\cite{pareek2025learningpowerflowconfidence,MOLLAALI2025107809,
REN24,Renkema2024PVForecast}. Conformal prediction is attractive because it
uses past differences between predictions and observations to construct
uncertainty intervals without assuming a specific probability distribution.
However, these intervals remain reliable only when past observations are
representative of the setting being assessed. This condition may fail after a new control policy is deployed at one or more inverters because this new policy can produce different closed-loop grid behaviour through its
interactions with the grid and the remaining controllers. Consequently,
uncertainty intervals calibrated under the pre-existing control policy may not
remain valid after deploying a new policy. 

This is an off-policy conformal prediction setting, in which data collected
under the existing control policies are used to assess a new one before deployment. Although off-policy conformal methods have been studied in
reinforcement-learning and multi-agent applications \cite{TAU22,KUI24}, to the
best of our knowledge, this work is the first in the power-systems literature
to formulate control policy replacement as an off-policy conformal problem for
grid-wide voltage-safety screening.

We propose \emph{Distributionally Robust Conformal Safety Screening}
(DR-CSS), a policy-agnostic statistical framework for pre-deployment,
scenario-by-scenario screening of a new control policy using historical data
and an imperfect nominal simulator. DR-CSS complements existing validation and verification tools and can be applied
to predictions produced by power-flow simulators, reduced-order models,
learned surrogates, or verification workflows. For each operating
scenario, the nominal simulator produces a future voltage trajectory,
around which DR-CSS constructs an uncertainty interval using historical
simulation-to-reality errors. The interval is then enlarged to account for
bounded changes in prediction-error behaviour after a new control policy is deployed at one or several inverters.

We further extend DR-CSS in two practical ways. First, context-aware
calibration adapts the uncertainty intervals to different operating
conditions, so that one overly conservative interval is not used for every
scenario. Second, gradual policy change introduces the new
control policy in stages and recalibrates DR-CSS after each stage, reducing the
change that must be handled if several inverters get a new control policy at once.

In summary, the main contributions are:
\begin{itemize}

     \item We introduce, to the best of our knowledge, the first framework in power systems for pre-deployment, scenario-by-scenario screening of a new control policy. \emph{Distributionally Robust Conformal Safety Screening} (DR-CSS) is
    model-agnostic and thus can be applied to any black-box control policy.

    \item We extend DR-CSS with context-aware calibration, which adapts the
    uncertainty intervals to different operating conditions, and gradual
    control policyreplacement, which updates the screening method after each
    deployment stage.

    \item Employing the IEEE 33-bus and 141-bus cases, we demonstrate how DR-CSS reaches the target statistical coverage with finite and informative uncertainty intervals, reduces the intervals' average width by up to \(69\%\), and
    flags every unsafe voltage trajectory in the gradual-deployment experiment.
\end{itemize}

The remainder of the paper is organized as follows. Section~II formulates the voltage-control and deployment-assessment problem. Section~III presents DR-CSS. Section~IV reports the numerical results. Finally, Section~V concludes the paper and outlines future work directions.


\section{Safety-Screening Problem Formulation}
\label{sec:problem}
Before introducing DR-CSS, this section defines the closed-loop
distribution-grid model, the joint control policy configuration, resulting from the combination of pre-existing control policies and new policies to be deployed, and
the scenario-level voltage-safety event assessed by the proposed method.

\subsection{Distributed Voltage-Control Setting}

Let $\mathcal{B}=\{1,\ldots,B\}$ denote the set of buses in a distribution grid, and let $\mathcal{I}_{\mathrm{inv}}\subseteq\mathcal{B}$ denote the subset of buses equipped with controllable converter-based reactive-power support and $\mathcal{L}\subseteq\mathcal{B}$ denote the set of buses equipped with loads. At time $t$ let
\(
    q_t := (Q_{i,t})_{i\in\mathcal{I}_{\mathrm{inv}}}
\)
collect the inverter reactive-power commands, and let $d_t$ denote uncontrollable operating conditions such as loads and active-power injections. The bus-voltage magnitudes are given by 
\(
    v_t := (V_{i,t})_{i\in\mathcal{B}},
\)
and $V_{\min}$ and $V_{\max}$ denote the minimum and maximum admissible voltage magnitudes, respectively.

We observe the grid state up to time $H$ and assess its future behavior over the look-ahead horizon
\(
    \mathcal{T}_{\mathrm{f}}:=\{H+1,\ldots,T\}, T>H .
\)
The main control target, defined by the Volt-Var problem, is to minimize reactive-power use while maintaining admissible bus voltages. If the future operating conditions were known and the inverter commands could be chosen directly at each time step, the problem could be formulated as follows:
\begin{subequations}
\label{eq:tv_opf}
\begin{align}
    \min_{\{q_t\}_{t\in\mathcal{T}_{\mathrm{f}}}}\quad &
    \sum_{t\in\mathcal{T}_{\mathrm{f}}}
    \sum_{i\in\mathcal{I}_{\mathrm{inv}}} Q_{i,t}^2
    \label{eq:tv_opf_obj}\\
    \text{s.t.}\quad &
    v_t=h_{\mathrm{pf}}(q_t,d_t),
    \quad \forall\,t\in\mathcal{T}_{\mathrm{f}},
    \label{eq:tv_opf_pf}\\
    &
    V_{\min}\le V_{i,t}\le V_{\max},
    \quad \forall\,i\in\mathcal{B},\; t\in\mathcal{T}_{\mathrm{f}},
    \label{eq:tv_opf_vlim}\\
    &
    |Q_{i,t}|\le Q_{i,t}^{\max}(P_{i,t}),
    \quad \forall\,i\in\mathcal{I}_{\mathrm{inv}},\; t\in\mathcal{T}_{\mathrm{f}} .
    \label{eq:tv_opf_qlim}
\end{align}
\end{subequations}
Here, $h_{\mathrm{pf}}$ is the AC power-flow mapping, $P_{i,t}$ is the active-power injection at inverter bus $i$, and the limit in \eqref{eq:tv_opf_qlim} is determined by the inverter apparent-power capability curve and based on the current output power of each inverter. The objective penalizes reactive-power usage, while constraints require all bus voltages to remain inside the admissible voltage band.

Please note that in this work, we do not discuss the implementation of the problem formulated in~\eqref{eq:tv_opf} nor do we develop a new control policy that employs it. Instead, we assume that a control policy is already specified and we study whether it can be safely deployed using historical operating data and a nominal simulator \(\Phi_{\mathcal M}\), such as an AC power-flow model, a reduced-order system, or a learned surrogate.

In reality, each inverter observes only local or neighboring measurements. Let \(X_t\) denote the physical system state, which may include voltages, nodal injections, forecasts, and other variables needed to describe the evolution of the closed-loop system, consisting of the distribution grid and the local control policies. The control policy need not observe \(X_t\) directly, but can rather have access to only a local observation $o_t$. For inverter \(i\), let \(\mathcal{N}_i\subseteq\mathcal{B}\) denote the set of buses whose voltage magnitudes are available to it, and define
\begin{equation}
\label{eq:local_observation}
o_{i,t}
:=
 h_i^{\mathrm{meas}}(X_t)
=
 \bigl(V_{j,t}\bigr)_{j\in\mathcal{N}_i}.
\end{equation}
The case \(\mathcal{N}_i=\{i\}\) corresponds to a local droop-type control policy, while larger neighborhoods allow the use of nearby measurements. Other inputs, such as active-power injections or forecasts, can be included through \(h_i^{\mathrm{meas}}\). Each inverter applies a control policy that outputs reactive power setpoint
\(
    Q_{i,t}=\pi_i(o_{i,t}),
\)
where \(\pi_i\) may be rule-based, optimization-based, or learning-based. The joint control policy, comprising all local policies in the grid, is denoted by
\(
    \pi := (\pi_i)_{i\in\mathcal{I}_{\mathrm{inv}}}.
\)
The closed-loop system evolves as
\begin{equation}
    X_{t+1}=F(X_t,q_t,W_t),
\end{equation}
where $F$ denotes the one-step physical state-transition mapping of the grid,
including the grid, device, and controller interactions, and \(W_t\) collects exogenous effects such as load changes, PV generation, forecast errors, and modeling disturbances. Thus, future voltages depend jointly on operating conditions, control policyactions, and the physical grid response.

Since voltage safety is a horizon-wide property, we consider the full future trajectory
\begin{equation}
    v_{H+1:T}:=(v_{H+1},\ldots,v_T).
\end{equation}
The worst voltage-limit violation over the look-ahead horizon is
\begin{equation}
\label{eq:gdef}
\begin{split}
    g(v_{H+1:T})
    :=
    \max_{t\in\mathcal{T}_{\mathrm{f}}}
    \max_{i\in\mathcal{B}}
    \Big\{\,
        &\max\{V_{i,t}-V_{\max},\,0\}, \\
        &\max\{V_{\min}-V_{i,t},\,0\}
    \Big\}.
\end{split}
\end{equation} The trajectory $v_{H+1:T}$ is safe if and only if \(g(v_{H+1:T})=0\).

\subsection{Safety Screening Goal}

Consider a grid historically operated under an existing baseline control policy $\pi^{\mathrm{base}}$, from which historical operating records are available. A new candidate policy $\pi^{\mathrm{new}}$ is proposed for deployment across a subset of inverter buses $\mathcal{E}\subseteq\mathcal{I}_{\mathrm{inv}}$. Starting at time $H$, the joint deployed policy $\pi^{\mathrm{dep}}$ satisfies:
\begin{equation}
\label{eq:deployed_policy}
Q_{i,t} = \begin{cases}
\pi_i^{\mathrm{new}}(o_{i,t}), & i\in\mathcal{E},\\
\pi_i^{\mathrm{base}}(o_{i,t}), & i\in\mathcal{I}_{\mathrm{inv}}\setminus\mathcal{E}.
\end{cases}
\end{equation}
Although DR-CSS can also accommodate added inverter-control units, this paper
focuses on replacing control policies at existing inverters. Given the observed grid history $X_{1:H}$ and forecasts of future uncontrollable operating conditions $\hat d_{H+1:T}$, we define the tested operating scenario as the tuple $x := (X_{1:H},\hat d_{H+1:T},\pi^{\mathrm{dep}})$. A nominal closed-loop simulator $\Phi_{\mathcal M}$ produces a predicted future reference voltage trajectory:
\begin{equation}
\label{eq:nominal_prediction}
\hat v_{H+1:T} = \Phi_{\mathcal M}(x).
\end{equation}
This trajectory is considered \textit{nominal} because it reflects idealized model and forecast assumptions rather than exact physical conditions.
However, a discrepancy between simulation and reality is almost inevitable, which results in a distribution shift in the data.
In our setting, two such shifts arise, and they are distinct rather than reducible to one another. \textit{Simulation-to-reality shift} reflects the fidelity of the simulator to the true grid model, that is, the discrepancy between simulated $\hat v_{H+1:T}$ and physically realized voltages $v_{H+1:T}$ due to unmodeled interactions, forecast errors, and measurement noise. \textit{Policy-induced shift}, in contrast, reflects the mismatch between the existing policy $\pi^{\mathrm{base}}$, under which historical calibration data was collected, and the deployment policy $\pi^{\mathrm{dep}}$, whose closed-loop interactions with the grid alter the operating trajectory even under identical underlying conditions. This second shift would persist even under a perfect simulator, since it originates from the feedback control loop itself rather than from model inaccuracy. For the remainder of this work, we use \textit{distribution shift} to refer jointly to both sources unless otherwise specified.

The central problem addressed in this paper is therefore:

\begin{quote}

Given historical records under a baseline control policy, an observed system history, forecasts of future conditions, a candidate control policy, and a nominal closed-loop simulator, our goal is to construct a data-driven rule that assesses the safety of test scenarios under the candidate policy and flags the scenarios that are uncertain or unsafe. 

\end{quote}

\section{Distributionally Robust Conformal Safety Screening}
\label{sec:dr_csl}

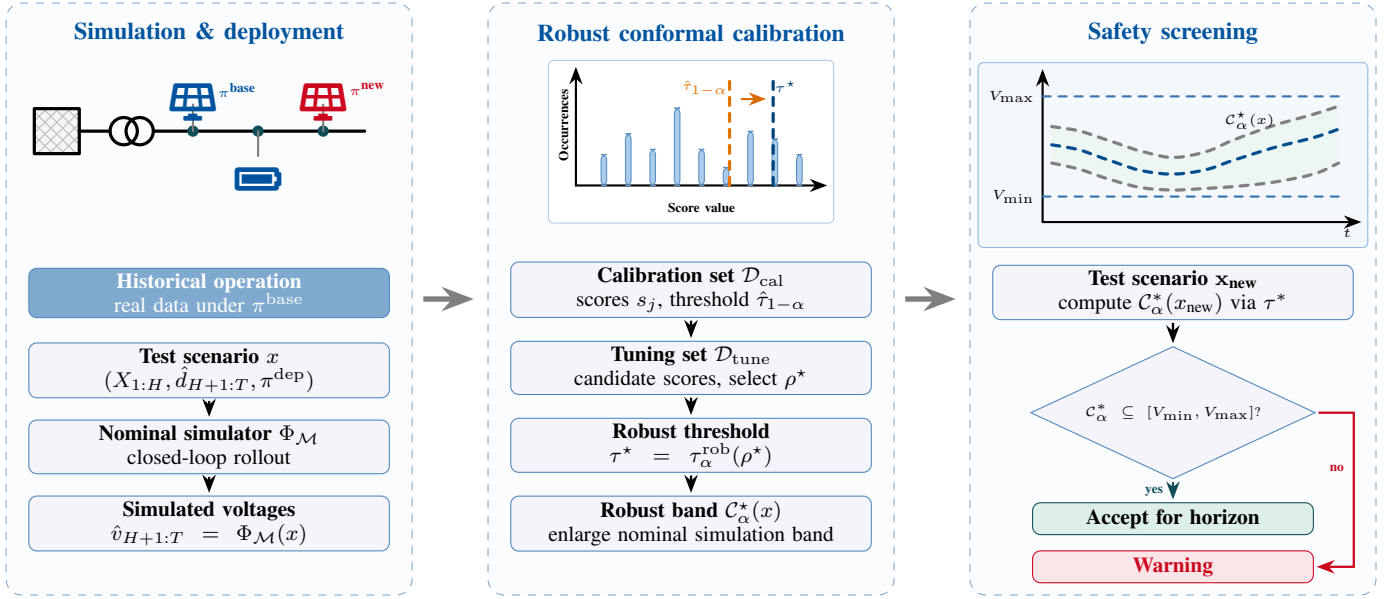
\begin{figure*}[t]
    \centering
     \begin{tikzpicture}[
    font=\rmfamily\scriptsize,
    >=Stealth,
    panel/.style={
        rectangle,
        draw,
        rounded corners=6pt,
        minimum width=5.35cm,
        minimum height=7.80cm,
        anchor=north,
        fill=white
    },
    block/.style={
        rectangle,
        draw,
        rounded corners=3pt,
        minimum width=4.75cm,
        text width=4.45cm,
        minimum height=0.58cm,
        inner ysep=2pt,
        align=center,
        font=\footnotesize,
        fill=white
    },
    greenbox/.style={block, draw=rwthpetrol100!80, fill=bglightgreen!30},
    purplebox/.style={block, draw=rwthblue100!70, fill=bglightpurple!40},
    bluebox/.style={block, draw=rwthblue100, fill=bglightblue!40},
    header/.style={align=center, font=\bfseries\small},
    decision/.style={
        diamond,
        draw=rwthblue100!70,
        fill=bglightpurple!40,
        aspect=2.15,
        inner sep=1pt,
        align=center,
        text width=3.05cm,
        minimum height=1.35cm,
        font=\tiny
    },
    resultbox/.style={
        block,
        minimum width=3.70cm,
        text width=3.40cm,
        minimum height=0.42cm,
        inner ysep=1pt
    }
]

\node[panel, draw=rwthblue100!50, dashed,  fill=bglightblue!15] (box1) at (0,0) {};
\node[panel, draw=rwthblue100!50, dashed, fill=bglightblue!15] (box2) at (6.35,0) {};
\node[panel, draw=rwthblue100!50, dashed, fill=bglightblue!15] (box3) at (12.70,0) {};

\node[header, text=rwthblue100] (h1) at ($(box1.north)+(0,-0.40)$)
{Simulation \& deployment};

\node[below=0.39cm of h1, inner sep=0pt] (gridfig) {
\begin{tikzpicture}[scale=0.78, line cap=round, line join=round]
    \begin{scope}[shift={(-2.75,0)}]
        \clip (-0.38,-0.38) rectangle (0.38,0.38);
        \fill[gray!8] (-0.38,-0.38) rectangle (0.38,0.38);
        \foreach \d in {-1.0,-0.8,...,1.0} {
            \draw[gray!55, thin] (-0.55,\d) -- (0.55,\d+1.10);
            \draw[gray!55, thin] (-0.55,\d+1.10) -- (0.55,\d);
        }
    \end{scope}
    \draw[very thick] (-3.13,-0.38) rectangle (-2.37,0.38);

    \draw[very thick] (-2.37,0) -- (-1.85,0);
    \draw[very thick, fill=none] (-1.62,0) circle (0.23);
    \draw[very thick, fill=none] (-1.35,0) circle (0.23);
    \draw[very thick] (-1.12,0) -- (2.45,0);

    \foreach \x in {-0.45,0.65,1.75} {
        \fill[rwthpetrol100] (\x,0) circle (0.095);
    }

    \draw[thick, rwthgray75] (-0.45,0) -- (-0.45,0.45);
    \node[font=\Large, text=rwthblue100] at (-0.45,0.82) {\faIcon{solar-panel}};
    \node[font=\tiny\bfseries, text=rwthblue100] at (0.3,0.82) {$\pi^{\text{base}}$};

    \draw[thick, rwthgray75] (0.65,0) -- (0.65,-0.42);
    \node[font=\Large, text=rwthblue100] at (0.65,-0.60) {\faIcon{battery-full}};

    \draw[thick, rwthgray75] (1.75,0) -- (1.75,0.45);
    \node[font=\Large, text=rwthred100] at (1.75,0.82) {\faIcon{solar-panel}};
    \node[font=\tiny\bfseries, text=rwthred100] at (2.5,0.82) {$\pi^{\text{new}}$};

\end{tikzpicture}
};

\node[
    purplebox,
    fill=rwthblue100!50,
    text=white,
    below=0.98cm of gridfig
] (hist)
{\textbf{Historical operation}\\real data under $\pi^{\mathrm{base}}$};

\node[purplebox, below=0.32cm of hist] (scen)
{\textbf{Test scenario $x$}\\$(X_{1:H},\hat d_{H+1:T},\pi^{\mathrm{dep}})$};

\node[purplebox, below=0.28cm of scen] (sim)
{\textbf{Nominal simulator $\Phi_{\mathcal M}$}\\closed-loop rollout};

\node[purplebox, below=0.28cm of sim] (volt)
{\textbf{Simulated voltages}\\$\hat v_{H+1:T}=\Phi_{\mathcal M}(x)$};


\node[header, text=rwthblue100] (h2) at ($(box2.north)+(0,-0.40)$)
{Robust conformal calibration};

\node[
    below=0.12cm of h2,
    inner sep=2pt,
    draw=rwthblue100!30,
    fill=white,
    rounded corners=2pt
] (plot) {
\begin{tikzpicture}[x=0.46cm, y=0.34cm, line cap=round, line join=round]
    \draw[->, thick] (0,0) -- (7.2,0);
    \draw[->, thick] (0,0) -- (0,4.6);

    \foreach \x/\h in {0.8/1.2,1.5/2.0,2.2/1.4,2.9/3.0,3.6/1.4,4.3/0.7,5.0/2.1,5.7/1.8,6.4/1.2} {
        \draw[fill=rwthblue50!75, draw=rwthblue100!65] (\x-0.08,0) rectangle (\x+0.08,\h);
    }

    \draw[dashed, very thick, orange!90!black] (4.4,0) -- (4.4,4.05);
    \draw[dashed, very thick, rwthblue100!80!black] (5.65,0) -- (5.65,4.05);
    \draw[->, thick, orange!85!black] (4.75,3.35) -- (5.45,3.35);

    \node[font=\tiny\bfseries, rotate=90] at (-0.65,2.35) {Occurrences};
    \node[font=\tiny\bfseries] at (3.6,-0.45) {Score value};
    \node[font=\tiny\bfseries, text=orange!90!black] at (3.7,4.30) {$\hat{\tau}_{1-\alpha}$};
    \node[font=\tiny\bfseries, text=rwthblue100!80!black] at (6.02,4.30) {$\tau^\star$};
\end{tikzpicture}
};

\node[purplebox, below=0.50cm of plot] (cal)
{\textbf{Calibration set $\mathcal D_{\mathrm{cal}}$}\\scores $s_j$, threshold $\hat{\tau}_{1-\alpha}$};

\node[purplebox, below=0.30cm of cal] (tun)
{\textbf{Tuning set $\mathcal D_{\mathrm{tune}}$}\\candidate scores, select $\rho^\star$};

\node[purplebox, below=0.30cm of tun] (thresh)
{\textbf{Robust threshold}\\$\tau^\star=\tau_{\alpha}^{\mathrm{rob}}(\rho^\star)$};

\node[purplebox, below=0.30cm of thresh] (band)
{\textbf{Robust band $\mathcal C_{\alpha}^{\star}(x)$}\\enlarge nominal simulation band};

\node[header, text=rwthblue100] (h3) at ($(box3.north)+(0,-0.40)$)
{Safety screening};

\node[
    below=0.12cm of h3,
    inner sep=2pt,
    draw=rwthblue50,
    fill=rwthblue50!10,
    rounded corners=2pt
] (plot2) {
\begin{tikzpicture}[x=0.60cm, y=0.48cm, line cap=round, line join=round]
    \fill[rwthblue50!10] (0,0) rectangle (6.8,4.0);

    \draw[->, thick] (0,0) -- (7.0,0);
    \draw[->, thick] (0,0) -- (0,4.2);

    \draw[dashed, thick, rwthblue100!80] (0,3.45) -- (6.55,3.45);
    \draw[dashed, thick, rwthblue100!80] (0,0.70) -- (6.55,0.70);

    \fill[bglightgreen!55]
        plot[smooth] coordinates {
            (0.20,2.62) (0.80,2.50) (1.45,2.24)
            (2.15,1.95) (2.85,1.78) (3.55,1.92)
            (4.25,2.28) (5.10,2.66) (5.95,2.96) (6.50,3.18)
        }
        -- plot[smooth] coordinates {
            (6.50,1.62) (5.95,1.32) (5.10,1.10)
            (4.25,0.98) (3.55,0.92) (2.85,0.88)
            (2.15,0.95) (1.45,1.18) (0.80,1.45) (0.20,1.62)
        }
        -- cycle;

    \draw[dashed, very thick, rwthgray75]
        plot[smooth] coordinates {
            (0.20,2.62) (0.80,2.50) (1.45,2.24)
            (2.15,1.95) (2.85,1.78) (3.55,1.92)
            (4.25,2.28) (5.10,2.66) (5.95,2.96) (6.50,3.18)
        };

    \draw[dashed, very thick, rwthgray75]
        plot[smooth] coordinates {
            (0.20,1.62) (0.80,1.45) (1.45,1.18)
            (2.15,0.95) (2.85,0.88) (3.55,0.92)
            (4.25,0.98) (5.10,1.10) (5.95,1.32) (6.50,1.62)
        };

    \draw[dashed, very thick, rwthblue100!90!black]
        plot[smooth] coordinates {
            (0.20,2.12) (0.80,1.98) (1.45,1.70)
            (2.15,1.42) (2.85,1.32) (3.55,1.43)
            (4.25,1.72) (5.10,2.05) (5.95,2.32) (6.50,2.55)
        };

    \node[font=\tiny\bfseries, anchor=east] at (-0.05,3.45) {$V_{\max}$};
    \node[font=\tiny\bfseries, anchor=east] at (-0.05,0.70) {$V_{\min}$};
    \node[font=\tiny\bfseries] at (4.55,3.18) {$\mathcal C_{\alpha}^{\star}(x)$};
    \node[font=\tiny\bfseries, anchor=north east] at (6.90,-0.02) {$t$};
\end{tikzpicture}
};

\node[purplebox, below=0.22cm of plot2] (new_scen)
{\textbf{Test scenario $\mathbf{x}_{\text{new}}$}\\compute $\mathcal{C}^{*}_{\alpha}(x_{\text{new}})$ via $\tau^*$};

\node[decision, below=0.34cm of new_scen] (check)
{$\mathcal{C}^*_{\alpha} \subseteq [V_{\min}, V_{\max}]$?};

\node[
    resultbox,
    draw=rwthpetrol100,
    fill=bglightgreen,
    below=0.3cm of check
] (accept)
{\textbf{Accept for horizon}};

\node[
    resultbox,
    draw=rwthred100,
    fill=rwthred100!10,
    below=0.2cm of accept
] (warn_wide)
{\textcolor{rwthred100}{\textbf{Warning}}};

\draw[->, line width=0.85pt, shorten >=1pt, shorten <=1pt]
    (scen.south) -- (sim.north);

\draw[->, line width=0.85pt, shorten >=1pt, shorten <=1pt]
    (sim.south) -- (volt.north);

\draw[->, line width=0.85pt, shorten >=1pt, shorten <=1pt]
    (cal.south) -- (tun.north);

\draw[->, line width=0.85pt, shorten >=1pt, shorten <=1pt]
    (tun.south) -- (thresh.north);

\draw[->, line width=0.85pt, shorten >=1pt, shorten <=1pt]
    (thresh.south) -- (band.north);

\draw[->, line width=0.85pt, shorten >=1pt, shorten <=1pt]
    (new_scen.south) -- (check.north);

\draw[->, line width=0.85pt, draw=rwthpetrol100, shorten >=1pt, shorten <=1pt]
    (check.south) -- node[left, font=\tiny\bfseries, text=rwthpetrol100] {yes}
    (accept.north);

\draw[->, line width=0.85pt, draw=rwthred100, shorten >=1pt, shorten <=1pt]
    (check.east) -- ++(0.50,0)
    |- node[left, pos=0.18, font=\tiny\bfseries, text=rwthred100] {no}
    (warn_wide.east);

\draw[->, line width=2pt, draw=rwthgray75]
    ([xshift=0.14cm]box1.east) -- ([xshift=-0.18cm]box2.west);

\draw[->, line width=2pt, draw=rwthgray75]
    ([xshift=0.14cm]box2.east) -- ([xshift=-0.18cm]box3.west);

\end{tikzpicture}
     \caption{Overview of the proposed DR-CSS workflow for safety screening of learning-based control policies under policy change. Historical data are collected under a baseline policy, while deployment may use a new candidate control policy at one or more inverter buses, shifting closed-loop voltage behavior and simulation-to-reality error statistics.
     }
    \label{fig:dr_csl_workflow}
\end{figure*}
This section presents the workflow of DR-CSS \footnote{The code for the DR-CSS implementation is publicly available at \href{https://github.com/sarraBou/DRCSS}{https://github.com/sarraBou/DRCSS}.}. 
Given a nominal simulator \(\Phi_{\mathcal M}\), DR-CSS compares simulated and realized trajectories under baseline operation to quantify the simulation-to-reality error. Furthermore, it compares the baseline and candidate closed-loop simulations to account for the shift induced by changing the control policy at one or several inverters. It then enlarges the nominal voltage prediction band to account for both effects.
Figure~\ref{fig:dr_csl_workflow} summarizes the workflow. The left panel defines the scenario information and nominal closed-loop simulation, the middle panel converts historical simulation-to-reality errors into robust conformal scores, and the right panel applies the final voltage band to return a scenario-level accept/warning decision.

DR-CSS is an offline pre-deployment screening procedure rather than a
voltage control policy. It applies the same decision rule independently to
each scenario and does not modify the control policy actions. A warning
identifies a scenario that requires further testing, policy revision,
or exclusion from the intended operating envelope.

\subsection{Scenario-Level Nominal Simulation}
This subsection corresponds to the left panel of
Fig.~\ref{fig:dr_csl_workflow}. DR-CSS evaluates the control policy separately
for each operating scenario. The observed grid history, forecasts of
future uncontrollable injections, and control policy to be evaluated are
provided to the nominal simulator in \eqref{eq:nominal_prediction}, which
produces one future voltage trajectory for that scenario.

If this trajectory is used directly for safety assessment, the simulator is implicitly treated as exact. Equivalently, the predictive distribution is the point mass

\begin{equation}
\label{eq:nominal-predictive-law}
\mathbb{P}^{\mathrm{nom}}_{x}
=
\delta_{\hat{v}_{H+1:T}(x)}.
\end{equation}
where $\delta_z$ denotes the Dirac probability measure concentrated at
$z$. $\mathbb{P}^{\mathrm{nom}}_{x}$ assigns all probability to the single simulated trajectory. However, this is too optimistic when the simulator is affected by forecast errors, model mismatch, measurement noise, controller delays, or unmodeled dynamics. DR-CSS therefore replaces this point-mass view by a \emph{data-calibrated statistical abstraction}: an uncertainty band around the nominal voltage trajectory.

\subsection{Voltage Error Scores and Conformal Calibration}
The next step is to learn this band from data. To do so, we compare historical nominal simulations with their realized voltage trajectories and summarize each discrepancy by a scalar conformal score. The following subsection describes the calibration block in Fig.~\ref{fig:dr_csl_workflow}, which maps a historical dataset of grid trajectories down to a single, rigorous safety threshold using split conformal prediction \cite{ANG22}. First, we pair past historical predictions from the nominal simulator with the true, realized voltage profiles measured from the physical grid. Second, for each historical scenario, we compare these profiles to calculate a single scalar \textit{nonconformity score} that captures the worst-case prediction error across all buses and time steps. We then collect these scores and sort them in ascending order. Finally, based on the target miscoverage level $\alpha$, we extract a benchmark safety threshold from the high end of this sorted list. If future simulation errors match these historical error patterns, wrapping this single threshold around the nominal simulation solution provides a coverage guarantee for the \textit{entire true physical voltage trajectory}, across all parts of the grid and all future hours simultaneously. Under exchangeability/calibration validity, this trajectory remains bounded within the calculated safety bands with finite-sample marginal coverage of at least $1-\alpha$. Let

\begin{equation}
\label{eq:D_cal}
\mathcal D_{\mathrm{cal}}
=
\left\{
\left(x^j,v^j_{H+1:T}\right)
\right\}_{j=1}^{n_{\mathrm{cal}}}.
\end{equation}

denote historical cases used for the calibration of the split conformal prediction method, where \(v^j_{H+1:T}\) is the realized voltage trajectory, and $n_{\mathrm{cal}}$ denotes the number of calibration samples. For each case, the simulator gives

$\hat v^j_{H+1:T}=\Phi_{\mathcal M}(x^j)$.
DR-CSS measures the simulation-to-reality discrepancy using the
trajectory-level standardized voltage-error score
\begin{equation}
\label{eq:score_trajectory}
s_j
:=
\max_{t\in\mathcal{T}_{\mathrm f}}
\max_{i\in\mathcal B}
\gamma_t
\frac{
\left|V_{i,t}^{j}-\widehat V_{i,t}^{j}\right|
}{
\widehat{\sigma}^{\mathrm{err}}_i
}.
\end{equation}
where $V_{i,t}^{j}$ is the realized voltage magnitude at bus $i$ and
time $t$ for calibration scenario $j$, and
$\widehat V_{i,t}^{j}$ is the corresponding nominal simulated voltage
magnitude, obtained from
$\widehat v^j_{H+1:T}=\Phi_{\mathcal M}(x^j)$.
Furthermore, $\widehat{\sigma}^{\mathrm{err}}_i>0$ is a fixed
bus-specific residual scale estimated before conformal calibration, and
$\gamma_t>0$ is a fixed weight for future time step $t$. The same values
of $\widehat{\sigma}^{\mathrm{err}}_i$ and $\gamma_t$ are used for the
calibration, tuning, and test scores. In the experiments, we use uniform
time weights, $\gamma_t=1$. Thus, $s_j$ is the largest standardized
voltage-prediction error across all buses and future time steps for
scenario $j$ and is dimensionless.

Let
\(
s_{(1)}\leq s_{(2)}\leq \cdots \leq s_{(n_{\mathrm{cal}})}
\)
be the sorted calibration scores. For a target miscoverage level \(\alpha\), define
\begin{equation}
\begin{aligned}
k_\alpha
&:=
\min\left\{
n_{\mathrm{cal}},
\left\lceil (n_{\mathrm{cal}}+1)(1-\alpha)\right\rceil
\right\}, \\
& \hat{\tau}_{1-\alpha}
:= s_{(k_\alpha)}
\end{aligned}
\end{equation}
Here, $\lceil a\rceil$ denotes the smallest integer greater than or equal
to $a$. For instance, \(\alpha=0.05\) corresponds to a nominal 95\% trajectory-level coverage target.
The conformal voltage interval is then

\begin{equation}
\label{eq:standard_voltage_interval}
\left|V_{i,t}-\widehat V_{i,t}\right|
\leq
\frac{
\widehat{\tau}_{1-\alpha}
\widehat{\sigma}^{\mathrm{err}}_i
}{
\gamma_t
},
\qquad
i\in\mathcal B,\;
t\in\mathcal{T}_{\mathrm f}.
\end{equation}
Thus, the simulated trajectory is surrounded by a voltage interval learned
from past simulation errors. Split conformal prediction guarantees the target coverage even with a
finite calibration dataset, provided that the calibration and future errors
are exchangeable, meaning that they follow the same underlying distribution. Here, the calibration errors are collected under the
baseline control policy; replacing it with the candidate policy may change
the closed-loop operating conditions and, consequently, the distribution
of future simulation errors.

\subsection{Robust Enlargement for Policy-Induced Error Changes}

The calibration scores defined in the previous section quantify simulation-to-reality error under baseline operation. Thus, the conformal interval defined above is reliable when future simulation errors behave like the calibration errors. However, changing the control policy at one or several inverters can additionally induce a policy-induced shift in the score distribution. As a result, the distribution of the score \(s\) may shift, and the historical simulation-error data may no longer be fully representative.

DR-CSS uses the calibration scores as historical examples of nominal-simulator error under baseline operation. Each score is a scalar: the largest weighted voltage prediction error over all buses and future time steps. Thus, DR-CSS does not model the full uncertain grid trajectory. Instead, it calibrates the distribution of this trajectory-level error score.
After control policy replacement, these scores may shift or develop a heavier tail because the new control policy can move the grid into different operating regimes. DR-CSS protects against this effect by replacing the standard conformal threshold with a higher score quantile. Here, \emph{distributionally robust} means robust to a bounded change in the scalar score distribution, not to a known full distribution of loads, PV generation, voltages, and control policy actions.
The robust step keeps the same split-conformal calibration scores, but uses a higher quantile than standard split conformal prediction. The parameter \(\rho\) controls this quantile inflation.  Larger values of \(\rho\) move the threshold to a higher calibration score and therefore produce a wider, more conservative voltage interval.

For the Total Variation (TV)-robust version, the inflated quantile level is
\begin{equation}   
\beta_{\alpha,\rho}^{\rm TV}
=
\min\left\{
1,\;
1-\alpha+\frac{\rho}{2}
\right\}.
\end{equation}
For a given robustness budget \(\rho\), define
\begin{equation}   
k_{\alpha,\rho}^{\rm TV}
=
\left\lceil
(n_{\rm cal}+1)\beta_{\alpha,\rho}^{\rm TV}
\right\rceil .
\end{equation}
If \(k_{\alpha,\rho}^{\rm TV}>n_{\rm cal}\), the calibration set does not
support a finite robust threshold at the requested confidence level, and
DR-CSS returns a warning. 
Otherwise, the robust threshold is
\begin{equation}
\label{eq:robust_threshold}
\tau_{\alpha}^{\mathrm{rob}}(\rho)
=
s_{\left(k_{\alpha,\rho}^{\mathrm{TV}}\right)}.
\end{equation}
Here, $s_{(k)}$ denotes the $k$-th order statistic, i.e., the $k$-th
smallest score among the calibration scores
$\{s_j\}_{j=1}^{n_{\mathrm{cal}}}$. The resulting robust voltage interval is
\begin{equation}
\label{eq:robust_voltage_interval}
\left|V_{i,t}-\widehat V_{i,t}\right|
\leq
\frac{
\tau_\alpha^{\mathrm{rob}}(\rho)
\widehat{\sigma}^{\mathrm{err}}_i
}{
\gamma_t
},
\qquad
i\in\mathcal B,\;
t\in\mathcal{T}_{\mathrm f}.
\end{equation}

Thus, $\rho$ does not directly modify the nominal simulated trajectory. Instead, it increases the score quantile used to construct the uncertainty band around that trajectory.

A small value of $\rho$ produces a relatively narrow interval but may provide insufficient protection against control policy-induced changes in simulation error. Conversely, a large value provides greater protection but may lead to unnecessarily conservative warning decisions. We therefore select $\rho$ before final testing using an independent tuning set generated under the candidate control policy $\pi^{\mathrm{dep}}$.

For a given robustness budget $\rho$, the empirical tuning coverage is
\begin{equation}
\label{eq:tuning_coverage}
\hat c_{\mathrm{tun}}(\rho) = \frac{1}{n_{\mathrm{tun}}} \sum_{m=1}^{n_{\mathrm{tun}}} \mathbf 1 \{s_m^{\mathrm{new}} \leq \tau_\alpha^{\mathrm{rob}}(\rho)\},
\end{equation}
where $s_m^{\mathrm{new}}$ is the simulation-error score for tuning case
$m$ under the new control policy. It is computed using
\eqref{eq:score_trajectory}, with the same fixed
$\widehat{\sigma}^{\mathrm{err}}_i$ and $\gamma_t$ used for calibration.

Because $\widehat c_{\mathrm{tun}}(\rho)$ is estimated from finitely many
tuning scenarios, it may overestimate the underlying coverage. We use the
one-sided Wilson lower confidence bound
\begin{align}
L_{\delta_{\mathrm{tun}}}(\rho)
&:=
\operatorname{WLB}
\left(
\widehat c_{\mathrm{tun}}(\rho),
n_{\mathrm{tun}},
\delta_{\mathrm{tun}}
\right),
\\
\operatorname{WLB}(\widehat p,n,\delta)
&:=
\max\left\{
0,\,
\frac{
\widehat p+\frac{a}{2}
-z\sqrt{
\frac{
\widehat p(1-\widehat p)+\frac{a}{4}
}{
n
}
}
}{
1+a
}
\right\},
\end{align}
where
\(
z=\Phi^{-1}(1-\delta),
a=\frac{z^2}{n}.
\) and \(\Phi\) is the standard normal cumulative distribution function.

The parameter $\delta_{\mathrm{tun}}$ specifies the error probability used
to construct the tuning-coverage confidence bound; the corresponding
confidence level is $1-\delta_{\mathrm{tun}}$. The lower confidence bound prevents the robustness budget from being selected solely on the basis of a potentially optimistic empirical coverage estimate.
The selected robustness budget is the smallest value whose lower confidence bound reaches the target coverage:
\begin{equation}
\label{eq:robustness_budget_selection}
\rho^\star = \inf \left\{ \rho \geq 0: L_{\delta_{\mathrm{tun}}}(\rho) \geq 1-\alpha \right\}.
\end{equation}

Because the robust threshold is non-decreasing in $\rho$, the empirical tuning coverage and its corresponding Wilson lower bound are also non-decreasing. This monotonicity enables an efficient bisection search for the smallest robustness budget satisfying the tuning criterion. Selecting the smallest feasible value limits unnecessary interval enlargement and reduces overly conservative warning decisions.

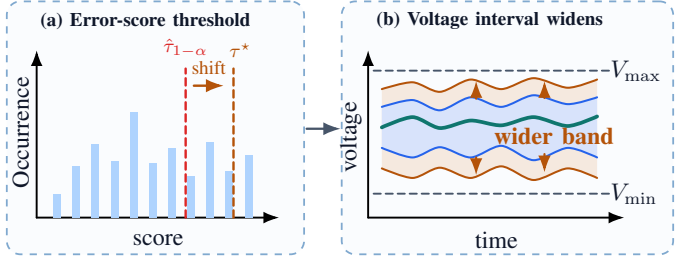
\begin{figure}[t]
    \centering
    \begin{tikzpicture}[
    font=\footnotesize,
    >=Latex,
    line cap=round,
    line join=round,
    axis/.style={line width=0.85pt, draw=cdark},
    flow/.style={-Latex, line width=0.95pt, draw=cgray},
    title/.style={font=\footnotesize\bfseries, text=cdark},
    tiny/.style={font=\rmfamily, text=cdark}
]

\begin{scope}[on background layer]
\node[rounded corners=2mm, draw=rwthblue100!50, dashed, fill=bglightblue!15, line width=0.75pt,
      minimum width=4.35cm, minimum height=3.70cm, anchor=south west] at (-0.45,-0.55) {};
\node[rounded corners=2mm, draw=rwthblue100!50, dashed, fill=bglightblue!15, line width=0.75pt,
      minimum width=4.85cm, minimum height=3.70cm, anchor=south west] at (4.45,-0.55) {};
\end{scope}

\node[title, anchor=west] at (-0.07,2.88) {(a) Error-score threshold};

\draw[->, thick] (0,0) -- (3.55,0);
\draw[->, thick] (0,0) -- (0,2.45);
\node[rotate=90, tiny] at (-0.23,1.25) {Occurrence};
\node[tiny] at (1.78,-0.32) {score};

\foreach \x/\h in {0.30/0.35,0.58/0.75,0.86/1.08,1.14/0.82,1.42/1.55,1.70/0.80,1.98/1.02,2.26/0.60,2.54/1.10,2.82/0.68,3.10/0.92}{
    \path[fill=cbar!75, draw=none] (\x-0.055,0) rectangle (\x+0.055,\h);
}

\draw[densely dashed, line width=1.0pt, draw=cred]    (2.18,0) -- (2.18,2.18);
\draw[densely dashed, line width=1.0pt, draw=corange] (2.88,0) -- (2.88,2.18);
\node[anchor=south, text=cred] at (2.18,2.23) {$\hat\tau_{1-\alpha}$};
\node[anchor=south, text=corange] at (2.98,2.23) {$\tau^\star$};

\draw[-Latex, line width=1.0pt, draw=corange] (2.30,1.93) -- (2.74,1.93);
\node[anchor=south, text=corange] at (2.52,1.97) {shift};

\draw[flow] (3.95,1.30) -- (4.45,1.30);

\begin{scope}[shift={(4.85,0)}]
\node[title, anchor=west] at (0,2.88) {(b) Voltage interval widens};

\draw[->, thick] (0,0) -- (3.75,0);
\draw[->, thick] (0,0) -- (0,2.45);
\node[rotate=90, tiny] at (-0.25,1.18) {voltage};
\node[tiny] at (1.88,-0.32) {time};

\draw[densely dashed, line width=0.8pt, draw=cgray] (0.08,2.15) -- (3.55,2.15);
\draw[densely dashed, line width=0.8pt, draw=cgray] (0.08,0.35) -- (3.55,0.35);
\node[anchor=west, tiny] at (3.42,2.15) {$V_{\max}$};
\node[anchor=west, tiny] at (3.42,0.35) {$V_{\min}$};

\fill[corange!13]
plot[smooth] coordinates {(0.20,1.88) (0.65,1.98) (1.05,1.84) (1.50,2.02) (1.95,1.88) (2.42,2.04) (2.92,1.90) (3.35,2.02)}
-- plot[smooth] coordinates {(3.35,0.58) (2.92,0.72) (2.42,0.55) (1.95,0.75) (1.50,0.58) (1.05,0.73) (0.65,0.58) (0.20,0.72)} -- cycle;
\draw[line width=0.85pt, draw=corange]
plot[smooth] coordinates {(0.20,1.88) (0.65,1.98) (1.05,1.84) (1.50,2.02) (1.95,1.88) (2.42,2.04) (2.92,1.90) (3.35,2.02)};
\draw[line width=0.85pt, draw=corange]
plot[smooth] coordinates {(0.20,0.72) (0.65,0.58) (1.05,0.73) (1.50,0.58) (1.95,0.75) (2.42,0.55) (2.92,0.72) (3.35,0.58)};

\fill[cblue!18]
plot[smooth] coordinates {(0.20,1.62) (0.65,1.72) (1.05,1.57) (1.50,1.76) (1.95,1.62) (2.42,1.79) (2.92,1.66) (3.35,1.76)}
-- plot[smooth] coordinates {(3.35,0.88) (2.92,1.02) (2.42,0.84) (1.95,1.05) (1.50,0.88) (1.05,1.03) (0.65,0.88) (0.20,1.02)} -- cycle;
\draw[line width=0.85pt, draw=cblue]
plot[smooth] coordinates {(0.20,1.62) (0.65,1.72) (1.05,1.57) (1.50,1.76) (1.95,1.62) (2.42,1.79) (2.92,1.66) (3.35,1.76)};
\draw[line width=0.85pt, draw=cblue]
plot[smooth] coordinates {(0.20,1.02) (0.65,0.88) (1.05,1.03) (1.50,0.88) (1.95,1.05) (2.42,0.84) (2.92,1.02) (3.35,0.88)};

\draw[line width=1.55pt, draw=cteal]
plot[smooth] coordinates {(0.20,1.32) (0.65,1.47) (1.05,1.31) (1.50,1.42) (1.95,1.35) (2.42,1.47) (2.92,1.34) (3.35,1.48)};

\draw[-Latex, line width=0.9pt, draw=corange] (1.58,1.74) -- (1.58,2.00);
\draw[-Latex, line width=0.9pt, draw=corange] (1.58,0.88) -- (1.58,0.62);
\draw[-Latex, line width=0.9pt, draw=corange] (2.58,1.75) -- (2.58,2.00);
\draw[-Latex, line width=0.9pt, draw=corange] (2.58,0.93) -- (2.58,0.66);
\node[anchor=west, text=corange, font=\rmfamily\bfseries] at (1.72,1.20) {wider band};

\end{scope}

\end{tikzpicture}
    \caption{Robust enlargement of the voltage uncertainty interval. Panel (a)
    shows the historical voltage error scores.
    Panel (b) shows the effect on voltages: the robust interval is wider
    than the standard one under control policy replacement.}
    \label{fig:robust_quantile}
\end{figure}

Once selected, $\rho^\star$ is fixed before evaluation on the independent test set. It is then used to construct the final robust voltage intervals shown in Fig.~\ref{fig:robust_quantile}. DR-CSS accepts the candidate deployment for an operating scenario only if the complete voltage band remains within the admissible limits over all buses and future time steps.

\subsection{Scenario-Level Accept/Warning Decision}
After calibration, the final robust interval is checked against the voltage limits for every bus and every time step in the horizon.
After selecting \(\rho^\star\), define
\(
\tau^\star := \tau_\alpha^{\mathrm{rob}}(\rho^\star)
\), as shown in Fig.~\ref{fig:dr_csl_workflow}.
For a tested scenario \(x_{\mathrm{new}}\), DR-CSS accepts the candidate deployment only if the full robust interval lies inside the voltage limits:
\begin{equation}
\label{eq:scenario_acceptance}
V_{\min}
\leq
\widehat V_{i,t}
-
\frac{
\tau^\star\widehat{\sigma}^{\mathrm{err}}_i
}{
\gamma_t
}
\quad\text{and}\quad
\widehat V_{i,t}
+
\frac{
\tau^\star\widehat{\sigma}^{\mathrm{err}}_i
}{
\gamma_t
}
\leq
V_{\max},
\end{equation}

for every \(i\in\mathcal B\) and \(t\in\mathcal{T}_{\mathrm f}\). Under the stated assumptions, the robust interval provides the prescribed
statistical coverage for the future voltage trajectory of the distribution grid
after deployment of the candidate control policy, jointly over all buses and
future time steps. Otherwise, DR-CSS issues a warning. A warning does not mean that a voltage violation will occur; it means that the nominal simulation, together with the calibration data and robustness budget, does not provide enough evidence to accept the deployment for that horizon.

\subsection{Context-Aware Calibration} \label{sec:context_aware}

The robust calibration introduced above uses a single threshold for all tested scenarios. However, the control policy-induced distribution shift depends on the operating condition, since changes in load demand and PV generation affect the inverter actions and the resulting voltage response. A global threshold may therefore be overly conservative in some conditions and insufficient in others.
To address this, we extend our method with a context-aware calibration. Specifically, we divide the scenarios into operating groups and apply
\eqref{eq:score_trajectory}-\eqref{eq:robustness_budget_selection}
separately within each group. Each scenario is assigned to a group using the grid net active power at the end of the observed history,
\(
P_{\mathrm{net}}
=
\sum_{i\in\mathcal{L}} P_i
-
\sum_{j\in\mathcal{I}_{\mathrm{inv}}} P_j\).
Negative values indicate PV-dominated operation, whereas larger values indicate higher net demand. The scenarios are divided into bins using empirical quantiles of \(P_{\mathrm{net}}\), and a separate robust threshold is calibrated for each bin. At test time, the threshold corresponding to the observed operating condition is used, while the accept/warning rule remains unchanged.
Net active power provides a simple measure of the balance between load demand and PV generation. Other variables, such as recent voltages, forecasts, or control policy-action differences, could also be used to group the scenarios; identifying the most informative context variable is left for future work.

\section{Experimental Results}
This section evaluates DR-CSS on the IEEE 33-bus and IEEE 141-bus systems
using droop control as the baseline and learning-based policies as the new
candidate policies, within single-control policy, context-aware, and gradual
replacement experiments.
\subsection{Experimental Setup}

\definecolor{ieeezoneone}{HTML}{00549F}
\definecolor{ieeezonetwo}{HTML}{57AB27}
\definecolor{ieeezonethree}{HTML}{A88500}
\definecolor{ieeezonefour}{HTML}{CC071E}
\definecolor{ieeepvorange}{HTML}{F6A800}
\definecolor{ieeegridline}{HTML}{4D4D4D}

\tikzset{
    ieee bus/.style={
        circle,
        draw=black,
        fill=white,
        line width=0.45pt,
        minimum size=3.6mm,
        inner sep=0pt,
        font=\tiny,
    },
    ieee zone label/.style={
        font=\scriptsize\bfseries,
        anchor=south,
    },
}

\newcommand{\drawieeepvpanelmark}[1]{
    \pgfscope
    \pgfsetstrokecolor{black}
    \pgfsetlinewidth{0.24pt}
    \pgfpathmoveto{\pgfpoint{0pt}{-.15\pgfplotmarksize}}
    \pgfpathlineto{\pgfpoint{0pt}{-.72\pgfplotmarksize}}
    \pgfpathmoveto{\pgfpoint{-.45\pgfplotmarksize}{-.72\pgfplotmarksize}}
    \pgfpathlineto{\pgfpoint{.45\pgfplotmarksize}{-.72\pgfplotmarksize}}
    \pgfusepathqstroke

    \pgfsetfillcolor{white}
    \pgfsetstrokecolor{#1}
    \pgfsetlinewidth{0.55pt}
    \pgfpathrectangle
        {\pgfpoint{-\pgfplotmarksize}{-.15\pgfplotmarksize}}
        {\pgfpoint{2\pgfplotmarksize}{1.0\pgfplotmarksize}}
    \pgfusepathqfillstroke

    \pgfsetstrokecolor{black}
    \pgfsetlinewidth{0.2pt}
    \pgfpathmoveto{\pgfpoint{-.34\pgfplotmarksize}{-.15\pgfplotmarksize}}
    \pgfpathlineto{\pgfpoint{-.34\pgfplotmarksize}{.85\pgfplotmarksize}}
    \pgfpathmoveto{\pgfpoint{.34\pgfplotmarksize}{-.15\pgfplotmarksize}}
    \pgfpathlineto{\pgfpoint{.34\pgfplotmarksize}{.85\pgfplotmarksize}}
    \pgfpathmoveto{\pgfpoint{-\pgfplotmarksize}{.35\pgfplotmarksize}}
    \pgfpathlineto{\pgfpoint{\pgfplotmarksize}{.35\pgfplotmarksize}}
    \pgfusepathqstroke
    \endpgfscope
}

\pgfdeclareplotmark{ieeepvpanel}{\drawieeepvpanelmark{black}}

\newcommand{\ieeepvunitbelow}[1]{%
    \begin{scope}[shift={($(#1.south)+(0,-2.9mm)$)}]
        \pgfsetplotmarksize{3.0pt}
        \pgfuseplotmark{ieeepvpanel}

        \coordinate (pvsuncenter) at (0,1.75mm);
        \fill[ieeepvorange] (pvsuncenter) circle[radius=0.70pt];
        \foreach \angle in {0,45,...,315}{
            \draw[ieeepvorange,line width=0.45pt]
                ($(pvsuncenter)+(\angle:1.10pt)$) --
                ($(pvsuncenter)+(\angle:1.90pt)$);
        }
    \end{scope}
}

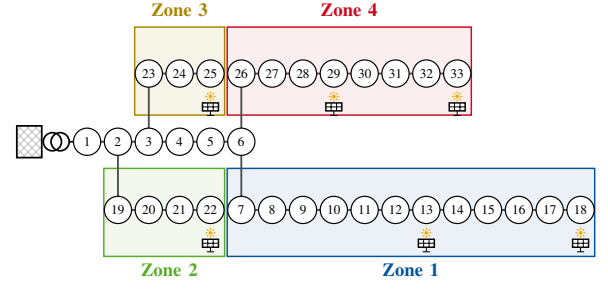
\begin{figure}[t]
\centering
\begin{tikzpicture}[x=0.34cm,y=0.82cm]
    \filldraw[
        fill=ieeezoneone!8,
        draw=ieeezoneone,
        line width=0.55pt,
    ] (5.45,-1.85) rectangle (19.75,-0.43);
    \filldraw[
        fill=ieeezonetwo!8,
        draw=ieeezonetwo,
        line width=0.55pt,
    ] (0.65,-1.85) rectangle (5.35,-0.43);
    \filldraw[
        fill=ieeezonethree!9,
        draw=ieeezonethree,
        line width=0.55pt,
    ] (1.85,0.43) rectangle (5.35,1.85);
    \filldraw[
        fill=ieeezonefour!7,
        draw=ieeezonefour,
        line width=0.55pt,
    ] (5.45,0.43) rectangle (14.95,1.85);

    \node[ieee zone label,anchor=north,text=ieeezoneone]
        at (12.60,-1.8) {Zone 1};
    \node[ieee zone label,anchor=north,text=ieeezonetwo]
        at (3.20, -1.8) {Zone 2};
    \node[ieee zone label,text=ieeezonethree]
        at (3.60,1.87) {Zone 3};
    \node[ieee zone label,text=ieeezonefour]
        at (10.20,1.87) {Zone 4};

    \coordinate (b1c) at (0,0);
    \coordinate (b2c) at (1.2,0);
    \coordinate (b3c) at (2.4,0);
    \coordinate (b4c) at (3.6,0);
    \coordinate (b5c) at (4.8,0);
    \coordinate (b6c) at (6.0,0);

    \foreach \bus/\x in {
        7/6.0,8/7.2,9/8.4,10/9.6,11/10.8,12/12.0,
        13/13.2,14/14.4,15/15.6,16/16.8,17/18.0,18/19.2
    }{
        \coordinate (b\bus c) at (\x,-1.1);
    }

    \foreach \bus/\x in {19/1.2,20/2.4,21/3.6,22/4.8}{
        \coordinate (b\bus c) at (\x,-1.1);
    }

    \foreach \bus/\x in {23/2.4,24/3.6,25/4.8}{
        \coordinate (b\bus c) at (\x,1.1);
    }

    \foreach \bus/\x in {
        26/6.0,27/7.2,28/8.4,29/9.6,
        30/10.8,31/12.0,32/13.2,33/14.4
    }{
        \coordinate (b\bus c) at (\x,1.1);
    }

    \draw[ieeegridline,line width=0.70pt]
        (b1c) -- (b2c) -- (b3c) -- (b4c) -- (b5c) -- (b6c);
    \draw[ieeegridline,line width=0.70pt]
        (b6c) -- (b7c) -- (b8c) -- (b9c) -- (b10c) -- (b11c)
        -- (b12c) -- (b13c) -- (b14c) -- (b15c) -- (b16c)
        -- (b17c) -- (b18c);
    \draw[ieeegridline,line width=0.70pt]
        (b2c) -- (b19c) -- (b20c) -- (b21c) -- (b22c);
    \draw[ieeegridline,line width=0.70pt]
        (b3c) -- (b23c) -- (b24c) -- (b25c);
    \draw[ieeegridline,line width=0.70pt]
        (b6c) -- (b26c) -- (b27c) -- (b28c) -- (b29c)
        -- (b30c) -- (b31c) -- (b32c) -- (b33c);

    \node[
        name=ieeeexternalgrid,
        draw=black,
        line width=0.65pt,
        rectangle,
        pattern=crosshatch,
        pattern color=gray!45,
        minimum width=3.3mm,
        minimum height=4.2mm,
        inner sep=0pt,
    ] at (-2.25,0) {};
    \node[
        name=ieeetrafoleft,
        draw=black,
        line width=0.65pt,
        circle,
        fill=none,
        minimum size=2.4mm,
        inner sep=0pt,
    ] at (-1.35,0) {};
    \node[
        name=ieeetraforight,
        draw=black,
        line width=0.65pt,
        circle,
        fill=none,
        minimum size=2.4mm,
        inner sep=0pt,
    ] at (-1.05,0) {};
    \draw[black,line width=0.65pt]
        (ieeeexternalgrid.east) -- (ieeetrafoleft.west);
    \draw[black,line width=0.65pt]
        (ieeetraforight.east) -- (b1c);

    \foreach \bus in {1,...,33}{
        \node[ieee bus] (b\bus) at (b\bus c) {\bus};
    }

    \ieeepvunitbelow{b13}
    \ieeepvunitbelow{b18}
    \ieeepvunitbelow{b22}
    \ieeepvunitbelow{b25}
    \ieeepvunitbelow{b29}
    \ieeepvunitbelow{b33}
\end{tikzpicture}
\caption{IEEE 33-bus feeder and the observation zones used by
the IDDPG actors.}
\label{fig:ieee33-pv-zones}
\end{figure}

We conduct our experiments using the open-source Multi-Agent Power
Distribution Network (MAPDN) environment~\cite{Wang2021}.
MAPDN provides PV generation and residential
load profiles, obtained from publicly available real-world datasets. The profiles are
resampled at a uniform temporal resolution of three minutes, which is
represented in the environment by the normalized time step
$\Delta t=1$. Each daily PV and load profile defines an operating scenario. Collectively, these scenarios cover a broad range of
generation and demand conditions. These profiles are assigned to two radial distribution-grid benchmarks: the IEEE 33-bus and IEEE 141-bus grids. The IEEE 33-bus and IEEE 141-bus radial grids contain
$6$ and $38$ PV-equipped inverter buses, respectively. 
Fig.~\ref{fig:ieee33-pv-zones} shows the IEEE 33-bus grid. In both systems, high PV penetration can
cause voltage-limit violations, motivating active voltage control.

For both grids, the admissible voltage range is set to 
    $[V_{\min},V_{\max}]
    =
    [0.95,\,1.05]\,\mathrm{p.u.}$
The reactive-power capability of inverter $i$ is constrained by its apparent-power rating, as defined in \eqref{eq:tv_opf_qlim}.
Consequently, the feasible reactive-power range changes dynamically
with the local active-power injection $P_{i,t}$. In this work we only consider reactive power control and assume that active power is an exogenous variable determined by the PV generation.\\
Trajectory windows are extracted from the daily PV and load profiles, resulting in distinct voltage trajectories.
The resulting trajectory windows are divided into disjoint training,
validation, calibration, and test sets \(2166\), \(334\), \(1500\), and \(1000\) samples,
respectively. These sets are used for predictor training, hyperparameter
tuning, conformal quantile estimation, and final evaluation, respectively. Each trajectory window consists of a 21-step observation history and a 10-step prediction horizon, corresponding to 63 minutes of history and a 30-minute look-ahead, at a resolution of 3 minutes. The look-ahead horizon is chosen based on the timescale of PV generation variability.

\subsubsection{Control Policies used in Experiments}

We consider two inverter-control configurations, a baseline droop control policy and a target RL control policy provided by the MAPDN benchmark~\cite{Wang2021}.

\begin{itemize}
\item Droop control policy: This policy implements a standard piecewise-linear Volt-Var
droop characteristic with a deadband. Reactive power is saturated outside the outer voltage limits, varies linearly in the droop regions, and is zero inside the deadband. Let \(V_{\mathrm{db}}^{-}=0.97 \text{ p.u.}\) and
\(V_{\mathrm{db}}^{+}=1.03 \text{ p.u.}\) denote the lower and upper deadband thresholds, the control rule can be defined as:
\begin{equation}
  Q_{i,t} =
  \begin{cases}
    Q_i^{\min},
      & V_{i,t} \leq V_{\min}, \\[2pt]
    Q_i^{\min}
    \dfrac{V_{\mathrm{db}}^{-}-V_{i,t}}
      {V_{\mathrm{db}}^{-}-V_{\min}},
      & V_{\min}<V_{i,t}<V_{\mathrm{db}}^{-}, \\[6pt]
    0,
      & V_{\mathrm{db}}^{-}\leq V_{i,t}\leq V_{\mathrm{db}}^{+}, \\[2pt]
    Q_i^{\max}
    \dfrac{V_{i,t}-V_{\mathrm{db}}^{+}}
      {V_{\max}-V_{\mathrm{db}}^{+}},
      & V_{\mathrm{db}}^{+}<V_{i,t}<V_{\max}, \\[6pt]
    Q_i^{\max},
      & V_{i,t} \geq V_{\max}.
  \end{cases}
  \label{eq:droop}
\end{equation}
Because the droop control depends only on local measurements, its
neighborhood is $\mathcal{N}_i=\{i\}$, and no communication with neighboring
buses is required.

\item MARL control policy: The second configuration employs the Independent Deep Deterministic Policy Gradient (IDDPG) algorithm implemented in the MAPDN
framework~\cite{Wang2021}. In this configuration, a subset of inverters $\mathcal{E}\subseteq\mathcal{I}_{\mathrm{inv}}$ is controlled by individual trained IDDPG actors. Each actor
observes the voltage magnitudes and active-power injections of the
inverter buses in its assigned zone $\mathcal N_i^\ast$. The zones for the
IEEE 33-bus case are shown in Fig.~3. We adopt the same zone definition as in the MAPDN framework. Each actor outputs a continuous reactive-power setpoint
\(
    Q_{i^{*}}(t)
    =
    \pi_{i^{*}}^{\mathrm{RL}}\!\left(o_{i^{*}}(t)\right)
    \in[Q^{\min}_{i^{*}},\,Q^{\max}_{i^{*}}].
\)
All other inverters $i \in\mathcal{I}_{\mathrm{inv}}\setminus\mathcal{E}$ continue to follow the droop control in
\eqref{eq:droop}. The IDDPG agents are trained using the L1-shaped voltage-barrier reward
introduced in~\cite{Wang2021}, and their actor and critic networks are trained in the
PandaPower-based MAPDN environment using the training scenarios, while
the droop-controlled buses remain fixed. After training, the RL actors, droop control policies, and simulator are fixed;
only the conformal calibration and robustness budget are estimated from the
calibration and tuning data.
\end{itemize}

\subsubsection{Benchmark Methods}
\label{sec:methods}

To demonstrate the efficiency of the proposed approach, we compare it to two benchmark methods. The compared methods use the same simulator $\Phi_{\mathcal{M}}$.
We set the conformal miscoverage level to
$\alpha_{\mathrm{CP}}=0.05$, giving \(95\%\) nominal coverage.

\begin{itemize}
    \item Standard conformal prediction (Standard CP):
    We apply split conformal prediction using the all-droop calibration
    set $\mathcal{D}_{\mathrm{cal}}$. The threshold
    $\hat{\tau}_{0.95}$ is obtained from the finite-sample-adjusted
    empirical quantile of the calibration scores $\{s_j\}$.

    \item Weighted MA-COPP:
    This state-of-the-art baseline applies weighted conformal
    prediction~\cite{TIB20} in the -agent -agent off-policy setting
    proposed in~\cite{KUI24}. Each calibration score is assigned the
    importance weight
    \(
        w_j
        =
        \frac{p_{\mathrm{test}}(x_j)}
             {p_{\mathrm{cal}}(x_j)}
    \).
    The density ratio is estimated by logistic regression using the
    terminal-history voltage and active-power features
    $(V_i(H),P_i(H))_{i\in\mathcal{B}}$. Under poor distributional overlap, large importance weights may cause the weighted
    critical value to be $+\infty$, yielding uninformative and unbounded intervals. As a fallback in such cases, we replace $+\infty$ with the largest admissible scalar score when constructing the voltage intervals. In this case, the original MA-COPP coverage guarantee no longer applies.
\end{itemize}
\subsection{Numerical Results}
For both IEEE systems, the following results consider a single-control policy replacement: one inverter uses the RL policy, while all others employ the droop control policy.

\subsubsection{Overall Performance}

\begin{table}[t]
\centering
\small
\renewcommand{\arraystretch}{1.}
\setlength{\tabcolsep}{3.2pt}
\caption{Average performance on the IEEE 33-bus and IEEE 141-bus cases.}
\label{tab:avg_performance_systems}

\begin{tabular*}{\columnwidth}{@{\extracolsep{\fill}}llcccc@{}}
\toprule
\textbf{System} & \textbf{Method}
& \textbf{Cov.}
& \textbf{Avg. width}
& \textbf{Max. width}
& \textbf{Unbd.} \\
&
& \textbf{95\%}
& \textbf{[p.u.]}
& \textbf{[p.u.]}
& \textbf{rate} \\
\midrule

\multirow{3}{*}{IEEE 33}
& CP
& \cellcolor{sRed!25} 0.946
& 0.0237
& 0.0498
& -- \\

& Weighted
& 0.999
& \cellcolor{sRed!25} 0.0791
& \cellcolor{sRed!25} 0.159
& \cellcolor{sRed!25} 1.000 \\

& DR-CSS
& \cellcolor{sGreen!25} \textbf{0.955}
& \cellcolor{sGreen!25} \textbf{0.0249}
& \cellcolor{sGreen!25} \textbf{0.0511}
& \cellcolor{sGreen!25} \textbf{0.000} \\

\midrule

\multirow{3}{*}{IEEE 141}
& CP
& \cellcolor{sRed!25} 0.945
& 0.040
& 0.054
& -- \\

& Weighted
& 1.000
& \cellcolor{sRed!25} 0.114
& \cellcolor{sRed!25} 0.157
& \cellcolor{sRed!25} 1.000 \\

& DR-CSS
& \cellcolor{sGreen!25} \textbf{0.965}
& \cellcolor{sGreen!25} \textbf{0.042}
& \cellcolor{sGreen!25} \textbf{0.058}
& \cellcolor{sGreen!25} \textbf{0.000} \\

\bottomrule
\end{tabular*}
\end{table}

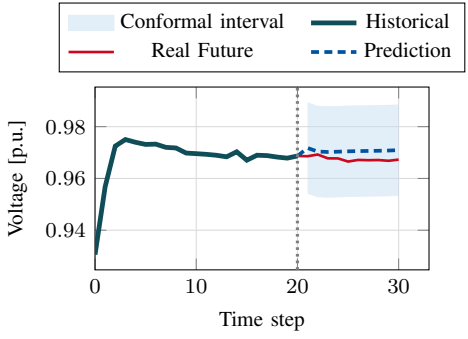
\begin{figure}[t]
    \centering
    \begin{tikzpicture}
        \begin{axis}[
            width=6cm, 
            height=4cm,
            xlabel={Time step},
            ylabel={Voltage [p.u.]},
            label style={font=\footnotesize},
            tick label style={font=\footnotesize},
            grid=both,
            grid style={line width=.1pt, draw=rwthgray75!30},
            legend columns=2,
            legend style={
                at={(0.5,1.08)},
                anchor=south,
                font=\footnotesize,
                draw=black,
                line width=0.4pt,
                fill=white,
                fill opacity=0.9,
                text opacity=1,
                /tikz/every even column/.append style={column sep=0.25cm}
            },
            unbounded coords=jump, 
            xmin=0,
            set layers=standard 
        ]
            
            \addplot[name path=A, transparent, forget plot] table [x=time, y=lower_interval, col sep=comma] {figures/results_ieee141/voltage_plot_data.csv};
            \addplot[name path=B, transparent, forget plot] table [x=time, y=upper_interval, col sep=comma] {figures/results_ieee141/voltage_plot_data.csv};
            \addplot[rwthblue50!50, opacity=0.5] fill between[of=A and B];
            \addlegendentry{Conformal interval}

            \addplot[color=rwthpetrol100, line width=1.8pt, no marks] table [x=time, y=historical, col sep=comma] {figures/results_ieee141/voltage_plot_data.csv};
            \addlegendentry{Historical}

            \addplot[color=rwthred100, line width=1.0pt, no marks] table [x=time, y=real_future, col sep=comma] {figures/results_ieee141/voltage_plot_data.csv};
            \addlegendentry{Real Future}

            \addplot[color=rwthblue100, line width=1.4pt, dash pattern=on 3pt off 2pt, no marks] table [x=time, y=prediction, col sep=comma] {figures/results_ieee141/voltage_plot_data.csv};
            \addlegendentry{Prediction}

            \draw[draw=rwthgray75, dash pattern=on 1pt off 1.5pt, line width=1.2pt] 
                (axis cs:20, \pgfkeysvalueof{/pgfplots/ymin}) -- (axis cs:20, \pgfkeysvalueof{/pgfplots/ymax});

        \end{axis}
    \end{tikzpicture}
    \caption{Voltage prediction with conformal intervals for one agent in the IEEE141 grid.}
        \label{fig:ieee141-voltage-example}
\end{figure}

Table~\ref{tab:avg_performance_systems} compares the empirical joint
trajectory coverage, average and maximum voltage-interval widths, and
unbounded-interval rate of the three methods. Standard CP attains an empirical
coverage of \(0.946\) and \(0.945\) on the IEEE 33 and IEEE 141, respectively, slightly below the nominal level
\(1-\alpha=0.95\). This undercoverage is consistent with a control policy-induced shift in the
nonconformity-score distribution: calibration errors are generated under
the all-droop control policy, whereas test errors arise after one inverter is
replaced by the RL control policy.
Weighted MA-COPP achieves near-perfect empirical coverage on both systems,
but all critical values are infinite, yielding unbounded intervals rate of \(100\%\).
We therefore report voltage interval widths obtained by capping the critical values at a finite scalar threshold. Thus, the MA-COPP coverage guarantee does not hold for the reported regions. 
In contrast, DR-CSS recovers the nominal coverage level with no unbounded intervals, while keeping the average interval width only \(3.0\%\) and \(5.0\%\) larger than Standard CP on the IEEE 33-bus and IEEE 141-bus systems, respectively. Compared with Weighted MA-COPP, the proposed robust approach reduces the average interval width by approximately \(69\%\) and \(63\%\) on the two systems. Overall,  DR-CSS compensates for the policy-induced shift with a modest efficiency cost, while the standard CP does not achieve the required coverage and MA-COPP does not provide the aimed guarantees.
Fig.~\ref{fig:ieee141-voltage-example} illustrates this effect for a representative voltage trajectory, where the conformal interval surrounds the nominal prediction over the forecast horizon and contains the realized future voltage.

To examine the spatial effect, Fig. \ref{fig:mean-interval-grid} shows the mean prediction-interval width for each bus in the IEEE 33-bus system. A clear spatial pattern emerges: prediction intervals become wider as the distance from the substation increases. Buses located near the beginning of a feeder have limited influence on the local voltage because
the substation has a dominant effect. Consequently, the baseline and new control policies induce similar voltage distributions, resulting in only a small distribution shift. Toward the end of the feeders, however, control policy actions have a greater influence on voltage. The replacement of the control policy, therefore, produces a more pronounced distribution shift, requiring wider prediction intervals. This observation highlights the importance of context: as discussed in Section \ref{sec:context_aware}, uncertainty depends on the spatial locations and the operating regime.

%

\providecommand{\gridpath}{figures/ablation_study}
\definecolor{gridline}{HTML}{666666}

\newcommand{\drawpvpanelmark}[1]{
    \pgfscope
    \pgfsetstrokecolor{black}
    \pgfsetlinewidth{0.24pt}
    \pgfpathmoveto{\pgfpoint{0pt}{-.15\pgfplotmarksize}}
    \pgfpathlineto{\pgfpoint{0pt}{-.72\pgfplotmarksize}}
    \pgfpathmoveto{\pgfpoint{-.45\pgfplotmarksize}{-.72\pgfplotmarksize}}
    \pgfpathlineto{\pgfpoint{.45\pgfplotmarksize}{-.72\pgfplotmarksize}}
    \pgfusepathqstroke

    \pgfsetfillcolor{white}
    \pgfsetstrokecolor{#1}
    \pgfsetlinewidth{0.55pt}
    \pgfpathrectangle
        {\pgfpoint{-\pgfplotmarksize}{-.15\pgfplotmarksize}}
        {\pgfpoint{2\pgfplotmarksize}{1.0\pgfplotmarksize}}
    \pgfusepathqfillstroke

    \pgfsetstrokecolor{black}
    \pgfsetlinewidth{0.2pt}
    \pgfpathmoveto{\pgfpoint{-.34\pgfplotmarksize}{-.15\pgfplotmarksize}}
    \pgfpathlineto{\pgfpoint{-.34\pgfplotmarksize}{.85\pgfplotmarksize}}
    \pgfpathmoveto{\pgfpoint{.34\pgfplotmarksize}{-.15\pgfplotmarksize}}
    \pgfpathlineto{\pgfpoint{.34\pgfplotmarksize}{.85\pgfplotmarksize}}
    \pgfpathmoveto{\pgfpoint{-\pgfplotmarksize}{.35\pgfplotmarksize}}
    \pgfpathlineto{\pgfpoint{\pgfplotmarksize}{.35\pgfplotmarksize}}
    \pgfusepathqstroke
    \endpgfscope
}

\pgfdeclareplotmark{pvpanel}{\drawpvpanelmark{black}}
\pgfplotsset{
    colormap={widthmap}{
        rgb255=(255,255,217);
        rgb255=(199,233,180);
        rgb255=(127,205,187);
        rgb255=(65,182,196);
        rgb255=(29,145,192);
        rgb255=(34,94,168);
        rgb255=(12,44,132)
    },
}

\pgfplotstableread[col sep=comma]{\gridpath/loadwise_rho_grid_meta.csv}\gridmeta
\pgfplotstablegetelem{0}{load_lower}\of{\gridmeta}
\edef\loadlower{\pgfplotsretval}
\pgfplotstablegetelem{0}{load_upper}\of{\gridmeta}
\edef\loadupper{\pgfplotsretval}
\pgfplotstablegetelem{0}{rho_uniform}\of{\gridmeta}
\edef\rhouniform{\pgfplotsretval}
\pgfmathparse{abs(\rhouniform) < 1e-9 ? 0 : \rhouniform}
\edef\rhouniform{\pgfmathresult}
\pgfplotstablegetelem{0}{q_uniform}\of{\gridmeta}
\edef\quniform{\pgfplotsretval}
\pgfplotstablegetelem{0}{width_min}\of{\gridmeta}
\edef\widthmin{\pgfplotsretval}
\pgfplotstablegetelem{0}{width_max}\of{\gridmeta}
\edef\widthmax{\pgfplotsretval}

\newcommand{\drawieeethirtythree}{
    \draw[gridline,line width=0.45pt]
        (axis cs:0,0) -- (axis cs:1.2,0) -- (axis cs:2.4,0)
        -- (axis cs:3.6,0) -- (axis cs:4.8,0) -- (axis cs:6.0,0);
    \draw[gridline,line width=0.45pt]
        (axis cs:6.0,0) -- (axis cs:6.0,-1.5) -- (axis cs:7.2,-1.5)
        -- (axis cs:8.4,-1.5) -- (axis cs:9.6,-1.5)
        -- (axis cs:10.8,-1.5) -- (axis cs:12.0,-1.5)
        -- (axis cs:13.2,-1.5) -- (axis cs:14.4,-1.5)
        -- (axis cs:15.6,-1.5) -- (axis cs:16.8,-1.5)
        -- (axis cs:18.0,-1.5) -- (axis cs:19.2,-1.5);
    \draw[gridline,line width=0.45pt]
        (axis cs:1.2,0) -- (axis cs:1.2,-1.5) -- (axis cs:2.4,-1.5)
        -- (axis cs:3.6,-1.5) -- (axis cs:4.8,-1.5);
    \draw[gridline,line width=0.45pt]
        (axis cs:2.4,0) -- (axis cs:2.4,1.5) -- (axis cs:3.6,1.5)
        -- (axis cs:4.8,1.5);
    \draw[gridline,line width=0.45pt]
        (axis cs:6.0,0) -- (axis cs:6.0,1.5) -- (axis cs:7.2,1.5)
        -- (axis cs:8.4,1.5) -- (axis cs:9.6,1.5)
        -- (axis cs:10.8,1.5) -- (axis cs:12.0,1.5)
        -- (axis cs:13.2,1.5) -- (axis cs:14.4,1.5);
    \draw[gridline,line width=0.35pt]
        (axis cs:13.2,-1.5) -- (axis cs:13.2,-2.05)
        (axis cs:19.2,-1.5) -- (axis cs:19.2,-2.05)
        (axis cs:4.8,-1.5) -- (axis cs:4.8,-2.05)
        (axis cs:4.8,1.5) -- (axis cs:4.8,0.95)
        (axis cs:9.6,1.5) -- (axis cs:9.6,0.95)
        (axis cs:14.4,1.5) -- (axis cs:14.4,0.95);
    \node[
        name=externalgrid,
        draw=black,
        line width=0.7pt,
        rectangle,
        pattern=crosshatch,
        pattern color=gray!45,
        minimum width=13pt,
        minimum height=13pt,
        inner sep=0pt
    ] at (axis cs:-2.55,0) {};
    \node[
        name=trafoleft,
        draw=black,
        line width=0.7pt,
        circle,
        fill=none,
        minimum size=8pt,
        inner sep=0pt
    ] at (axis cs:-1.40,0) {};
    \node[
        name=traforight,
        draw=black,
        line width=0.7pt,
        circle,
        fill=none,
        minimum size=8pt,
        inner sep=0pt
    ] at (axis cs:-1.05,0) {};
    \draw[gridline,line width=0.7pt]
        (externalgrid.east) -- (trafoleft.west);
    \draw[gridline,line width=0.7pt]
        (traforight.east) -- (axis cs:0,0);
}

\newcommand{\plotagentdata}[1]{
    \addplot[
        only marks,
        scatter,
        scatter src=explicit,
        scatter/use mapped color={draw=black,fill=mapped color},
        mark=*,
        mark size=2.25pt,
        point meta min=\widthmin,
        point meta max=\widthmax,
        restrict expr to domain={\thisrow{plot_class}}{0:1},
        unbounded coords=discard,
    ] table[
        x=plot_x,
        y=plot_y,
        meta=mean_width,
        col sep=comma
    ] {#1};

    \addplot[
        only marks,
        mark=pvpanel,
        mark size=3.0pt,
        restrict expr to domain={\thisrow{plot_class}}{2:3},
        unbounded coords=discard,
    ] table[
        x=plot_x,
        y=plot_y,
        col sep=comma
    ] {#1};
}

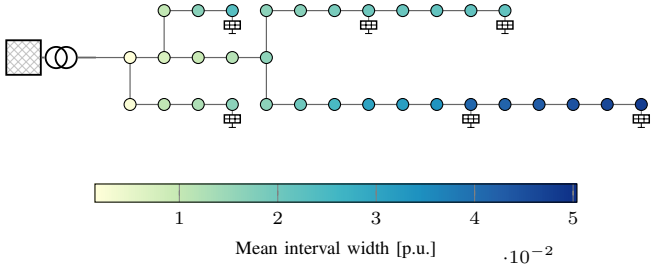
\begin{figure}[t]
\centering
\begin{tikzpicture}
\begin{axis}[
    name=uniformgrid,
    width=0.98\columnwidth,
    height=0.215\columnwidth,
    xmin=-3.1,
    xmax=20.0,
    ymin=-2.35,
    ymax=2.25,
    scale only axis,
    hide axis,
    clip=false,
    filter discard warning=false,
    colormap name=widthmap]
\drawieeethirtythree
\plotagentdata{\gridpath/loadwise_rho_grid_uniform.csv}
\end{axis}

\begin{axis}[
    name=widthcolorbar,
    at={($(uniformgrid.south)+(0,-4mm)$)},
    anchor=north,
    hide axis,
    scale only axis,
    width=0.72\columnwidth,
    height=0pt,
    point meta min=\widthmin,
    point meta max=\widthmax,
    colormap name=widthmap,
    colorbar horizontal,
    colorbar style={
        width=0.72\columnwidth,
        height=2.4mm,
        xlabel={Mean interval width [p.u.]},
        xlabel style={font=\scriptsize},
        xticklabel style={
            font=\scriptsize,
            /pgf/number format/fixed,
            /pgf/number format/precision=3,
        },
    },
]
\addplot[draw=none] coordinates {(0,0) (1,1)};
\end{axis}
\end{tikzpicture}
\caption{
Mean interval width for each bus in the IEEE33 case.
}
\label{fig:mean-interval-grid}
\end{figure}

\subsubsection{Context-Aware Calibration}

\providecommand{\threewaywidthdatapath}{figures/threeway_loadwise_width_boxplot}

\usepgfplotslibrary{statistics}
\pgfplotsset{compat=1.18}

\providecolor{rwthblue100}{RGB}{0,84,159}
\providecolor{rwthorange100}{HTML}{0F4D58}
\providecolor{rwthgray75}{RGB}{100,101,103}

\pgfplotstableread[
  col sep=comma
]{\threewaywidthdatapath/width_boxplot_stats.csv}\threewaywidthstats

\newcommand{\addthreewaywidthbox}[3]{%
  \pgfplotstablegetelem{#1}{whisker_low}\of{\threewaywidthstats}%
  \edef\boxlow{\pgfplotsretval}%
  \pgfplotstablegetelem{#1}{q1}\of{\threewaywidthstats}%
  \edef\boxqone{\pgfplotsretval}%
  \pgfplotstablegetelem{#1}{median}\of{\threewaywidthstats}%
  \edef\boxmedian{\pgfplotsretval}%
  \pgfplotstablegetelem{#1}{q3}\of{\threewaywidthstats}%
  \edef\boxqthree{\pgfplotsretval}%
  \pgfplotstablegetelem{#1}{whisker_high}\of{\threewaywidthstats}%
  \edef\boxhigh{\pgfplotsretval}%
  %
  \edef\preparedboxcommand{%
    \noexpand\addplot+[
      #3,
      forget plot,
      boxplot prepared={
        draw position=#2,
        lower whisker=\boxlow,
        lower quartile=\boxqone,
        median=\boxmedian,
        upper quartile=\boxqthree,
        upper whisker=\boxhigh
      }
    ] coordinates {};
  }%
  \preparedboxcommand
}

\begin{figure}[t]
\centering
\begin{tikzpicture}
\begin{axis}[
  width=5.8cm,
  height=4cm,
  xmin=0.5,
  xmax=5.5,
  xtick={1,2,3,4,5},
  xticklabels={1,2,3,4,5},
  xlabel={Load bin},
  ylabel={\shortstack{Prediction-interval\\width [p.u.]}},
  label style={font=\footnotesize},
  tick label style={font=\footnotesize},
  scaled y ticks=false,
  yticklabel style={
    font=\footnotesize,
    /pgf/number format/fixed,
    /pgf/number format/precision=2
  },
  grid=both,
  grid style={line width=.1pt,draw=rwthgray75!30},
  legend columns=2,
  legend style={
    at={(0.5,1.16)},
    anchor=south,
    font=\footnotesize,
    draw=black,
    line width=0.4pt,
    fill=white,
    fill opacity=0.9,
    text opacity=1,
    /tikz/every even column/.append style={column sep=0.05cm}
  },
  legend image code/.code={
    \draw[#1] (0cm,-0.08cm) rectangle (0.38cm,0.08cm);
  },
  unbounded coords=jump,
  set layers=standard,
  boxplot/every box/.style={solid,line width=0.8pt},
  boxplot/every whisker/.style={solid,line width=0.7pt},
  boxplot/every median/.style={black,line width=1.0pt},
  boxplot/every average/.style={mark=none},
  boxplot/box extend=0.27,
  boxplot/draw direction=y
]

\addlegendimage{
  area legend,
  draw=rwthblue100,
  fill=rwthblue100!24
}
\addlegendentry{Uniform DR-CSS}

\addlegendimage{
  area legend,
  draw=rwthorange100,
  fill=rwthorange100!24
}
\addlegendentry{Loadwise DR-CSS}

\addthreewaywidthbox{0}{0.82}{draw=rwthblue100,fill=rwthblue100!24}
\addthreewaywidthbox{1}{1.18}{draw=rwthorange100,fill=rwthorange100!24}

\addthreewaywidthbox{2}{1.82}{draw=rwthblue100,fill=rwthblue100!24}
\addthreewaywidthbox{3}{2.18}{draw=rwthorange100,fill=rwthorange100!24}

\addthreewaywidthbox{4}{2.82}{draw=rwthblue100,fill=rwthblue100!24}
\addthreewaywidthbox{5}{3.18}{draw=rwthorange100,fill=rwthorange100!24}

\addthreewaywidthbox{6}{3.82}{draw=rwthblue100,fill=rwthblue100!24}
\addthreewaywidthbox{7}{4.18}{draw=rwthorange100,fill=rwthorange100!24}

\addthreewaywidthbox{8}{4.82}{draw=rwthblue100,fill=rwthblue100!24}
\addthreewaywidthbox{9}{5.18}{draw=rwthorange100,fill=rwthorange100!24}

\end{axis}
\end{tikzpicture}

\caption{Prediction-interval width in each load bin for uniform and loadwise robust conformal prediction.}
\label{fig:threeway-loadwise-width-boxplot}
\end{figure}
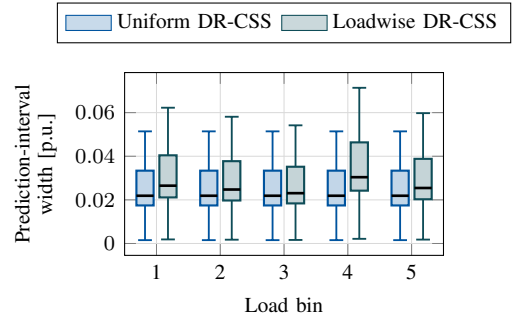

This section evaluates the context-aware calibration described in
Section~\ref{sec:context_aware}. We consider the IEEE 33-bus system and
replace the droop control policy at bus~25 with the RL control policy. This
corresponds to a control policy change at one inverter. The scenarios are divided into five bins using empirical quantiles of \(P_{\mathrm{net}}\) in the shifted simulation set, yielding 400 simulated trajectories per bin.

Table~\ref{tab:load_sensitivity_coverage} reports the empirical coverage of
standard CP, uniform DR-CSS with a single context-independent threshold, and loadwise DR-CSS with loadwise calibration. The results reveal high variation across operating regimes. For standard CP and uniform DR-CSS, the two most PV-dominated bins (bins 1 and 2) attain the target coverage of $1-\alpha=0.95$, whereas the other bins fall below the target, with the largest undercoverage occurring in bin~4.
The robust correction primarily accounts for the distribution shift induced by the change in control policy. Because control policy actions have a stronger effect on voltage during PV-dominated operating conditions, this correction may be more conservative in those regimes. In load-dominated conditions, however, the remaining uncertainty may be governed by operating-regime effects not fully captured by the control policy-shift correction, leading to larger conditional nonconformity scores and undercoverage. More generally, neither standard CP nor uniform DR-CSS guarantees coverage within each load bin: attaining the target marginally over the full dataset does not imply valid coverage within individual subgroups of the data. The smaller number of observations within each bin may also contribute to the observed variability.

Loadwise calibration, on the other hand, accounts for this regime dependence and attains the coverage target in every bin. This improvement comes at the cost of
moderately wider intervals, as shown in
Fig.~\ref{fig:threeway-loadwise-width-boxplot}. The increase is most pronounced in bin~4, where the uniform methods exhibit the greatest undercoverage, indicating that this regime requires a larger conditional critical value.

\begin{table}[t]
\centering
\caption{Comparison of the empirical coverage of standard CP, uniform DR-CSS and loadwise DR-CSS across
operating-point bins defined by \(P_{\mathrm{net}}\). }
\label{tab:load_sensitivity_coverage}
\begin{tabular}{lccccc}

\toprule
\textbf{Bin} 
& \textbf{\(P_{\mathrm{net}}\) range }
& \textbf{Standard}
& \textbf{Uniform}
& \textbf{Loadwise} \\

\textbf{Index}
& \textbf{[p.u.]}
& \textbf{CP Cov.}
& \textbf{DR-CSS Cov.}
& \textbf{DR-CSS Cov.}\\
\midrule

1 & \([-5.86,\,-3.32)\) &  0.980 & 0.985 & 0.995  \\
2 & \([-3.32,\,-2.48)\) & 0.962 & 0.962 & 0.967 \\
3 & \([-2.48,\,-1.68)\) & \cellcolor{sRed!25} 0.936 & \cellcolor{sRed!25} 0.948 & 0.953 \\
4 & \([-1.68,\,-0.62)\) & \cellcolor{sRed!25} 0.921 & \cellcolor{sRed!25} 0.942 & 0.995 \\
5 & \([-0.62,\, 1.96]\) & \cellcolor{sRed!25} 0.926 & \cellcolor{sRed!25} 0.944 & 0.974\\
\bottomrule
\end{tabular}
\end{table}

\subsubsection{Abrupt and Gradual Control Policy Replacement}

The previous experiments considered the replacement of the control policy at a single inverter bus. In practice, however, the control policy at multiple inverters may be updated during deployment. When several control policies are replaced at once, the shift in the score distribution can become larger. This requires a stronger robustness correction, which makes the screening more conservative and can increase the number of warning decisions. We therefore propose and evaluate a gradual deployment strategy in which control policies are introduced one at a time. After each step, new data under the new joint policy are collected and the calibration and tuning procedure is repeated. This reduces the shift handled at each stage and can lower unnecessary warnings while preserving the same accept/warning rule.

The previous experiments considered a single policy change at bus 25. Now, we consider the control policy replacement at buses 25, 29, and 13. In the \emph{abrupt} protocol, all three inverters switch to RL control simultaneously, so calibration starts from the all-droop control and targets the three-RL-control configuration (3RL). In the \emph{gradual} strategy, inverters are
replaced sequentially \(25\rightarrow29\rightarrow13\); after each replacement, new trajectories are collected and the conformal method is recalibrated. For comparability, both strategies are
evaluated on the same real $3\mathrm{RL}$ deployment dataset.

\providecommand{\datapath}{figures/gradual_vs_abrupt}

\definecolor{rwthblue100}{HTML}{00549F}
\definecolor{rwthpetrol100}{HTML}{0F4D58}
\definecolor{rwthred100}{HTML}{CC071E}

\pgfplotstableread[col sep=comma]{\datapath/2_q_change.csv}\qdata

\pgfplotstablegetelem{0}{q_cp}\of{\qdata}\edef\abruptqcp{\pgfplotsretval}
\pgfplotstablegetelem{0}{q_star}\of{\qdata}\edef\abruptqstar{\pgfplotsretval}
\pgfplotstablegetelem{0}{tax}\of{\qdata}\edef\abrupttax{\pgfplotsretval}
\pgfplotstablegetelem{1}{q_cp}\of{\qdata}\edef\gradualqcp{\pgfplotsretval}
\pgfplotstablegetelem{1}{q_star}\of{\qdata}\edef\gradualqstar{\pgfplotsretval}
\pgfplotstablegetelem{1}{tax}\of{\qdata}\edef\gradualtax{\pgfplotsretval}

\pgfmathsetmacro{\abrupttaxx}{0.5*(\abruptqcp+\abruptqstar)}
\pgfmathsetmacro{\gradualtaxx}{0.5*(\gradualqcp+\gradualqstar)}

\pgfplotstableread[col sep=comma]{\datapath/4_coverage_vs_width.csv}\covwidthdata

\pgfplotstablegetelem{0}{coverage_mean}\of{\covwidthdata}\edef\abruptcoverage{\pgfplotsretval}
\pgfplotstablegetelem{0}{width_mean}\of{\covwidthdata}\edef\abruptwidth{\pgfplotsretval}
\pgfplotstablegetelem{0}{target_cov}\of{\covwidthdata}\edef\nominalcoverage{\pgfplotsretval}
\pgfplotstablegetelem{1}{coverage_mean}\of{\covwidthdata}\edef\gradualcoverage{\pgfplotsretval}
\pgfplotstablegetelem{1}{width_mean}\of{\covwidthdata}\edef\gradualwidth{\pgfplotsretval}

\pgfmathsetmacro{\widthymax}{1.18*max(\abruptwidth,\gradualwidth)}

\begin{figure}[t]
\centering
\begin{tikzpicture}
\begin{groupplot}[
    group style={
        group size=2 by 1,
        horizontal sep=14mm,
    },
    width=0.295\columnwidth,
    height=0.34\columnwidth,
    scale only axis,
    tick label style={font=\scriptsize},
    label style={font=\scriptsize},
    axis line style={black!65},
    tick style={black!65},
    clip=false,
]

\nextgroupplot[
    xlabel={Threshold $\tau$},
    ytick={1,0},
    yticklabels={Abrupt,Gradual},
    xmin=3.5,
    xmax=12.5,
    ymin=-0.45,
    ymax=1.42,
    xtick={4,8,12},
]
\addplot[rwthred100, very thick]
    coordinates {(\abruptqcp,1) (\abruptqstar,1)};

\addplot[rwthpetrol100, very thick]
    coordinates {(\gradualqcp,0) (\gradualqstar,0)};

\addplot[
    only marks,
    mark=*,
    mark size=2.4pt,
    draw=rwthred100,
    fill=white,
    line width=0.8pt,
] coordinates {(\abruptqcp,1)};

\addplot[
    only marks,
    mark=*,
    mark size=2.6pt,
    draw=black,
    fill=rwthred100,
] coordinates {(\abruptqstar,1)};

\addplot[
    only marks,
    mark=*,
    mark size=2.4pt,
    draw=rwthpetrol100,
    fill=white,
    line width=0.8pt,
] coordinates {(\gradualqcp,0)};

\addplot[
    only marks,
    mark=*,
    mark size=2.6pt,
    draw=black,
    fill=rwthpetrol100,
] coordinates {(\gradualqstar,0)};

\node[
    font=\scriptsize,
    anchor=south,
    text=rwthred100,
    fill=white,
    fill opacity=0.92,
    text opacity=1,
    inner sep=1pt,
]
at (axis cs:\abrupttaxx,1.08)
{$+\pgfmathprintnumber[fixed,precision=2]{\abrupttax}$};

\node[
    font=\scriptsize,
    anchor=south,
    text=rwthpetrol100,
    fill=white,
    fill opacity=0.92,
    text opacity=1,
    inner sep=1pt,
]
at (axis cs:\gradualtaxx,0.08)
{$+\pgfmathprintnumber[fixed,precision=2]{\gradualtax}$};

\nextgroupplot[
    xlabel={Coverage},
    ylabel={Width (p.u.)},
    xmin=0.94,
    xmax=1.003,
    ymin=0,
    ymax=\widthymax,
    xtick={0.95,0.98,1.00},
    ytick={0,0.03,0.06},
    scaled y ticks=false,
    y tick label style={
        /pgf/number format/fixed,
        /pgf/number format/precision=2,
        font=\scriptsize,
    },
    x tick label style={font=\scriptsize},
]
\addplot[
    only marks,
    mark=*,
    mark size=3.4pt,
    draw=black,
    fill=rwthred100,
] coordinates {(\abruptcoverage,\abruptwidth)};

\addplot[
    only marks,
    mark=*,
    mark size=3.4pt,
    draw=black,
    fill=rwthpetrol100,
] coordinates {(\gradualcoverage,\gradualwidth)};

\addplot[sharp plot, no marks, rwthblue100, dashed, thick]
    coordinates {(\nominalcoverage,0) (\nominalcoverage,\widthymax)};

\node[font=\scriptsize, anchor=south east, xshift=-1pt, yshift=1pt]
    at (axis cs:\abruptcoverage,\abruptwidth) {Abrupt};

\node[font=\scriptsize, anchor=north east, xshift=-1pt, yshift=-1pt]
    at (axis cs:\gradualcoverage,\gradualwidth) {Gradual};

\end{groupplot}

\node[font=\footnotesize, anchor=north west, align=left]
    at ($(group c1r1.south west)+(0,-0.75cm)$)
    {(a) Threshold};

\node[font=\footnotesize, anchor=north west, align=left]
    at ($(group c2r1.south west)+(0,-0.75cm)$)
    {(b) Coverage--width};
\end{tikzpicture}
\caption{Comparison of abrupt and gradual distribution shifts. Open markers in panel~(a) denote the standard CP threshold and filled markers the robust threshold; panel~(b) reports the resulting empirical coverage and mean interval width.}
\label{fig:gradual-abrupt-combined}
\end{figure}
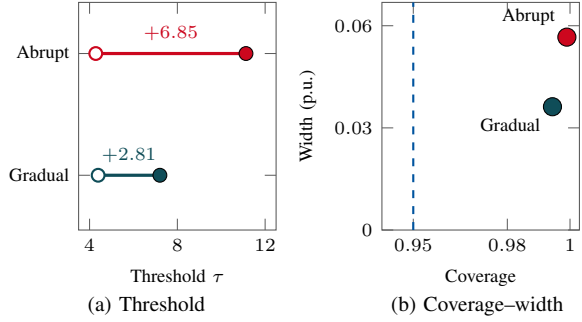

Figures~\ref{fig:gradual-abrupt-combined} (a)-(b) show that the proposed gradual deployment strategy reduces conservativeness while maintaining empirical coverage above the nominal target under the same final 3RL deployment. In Fig.~\ref{fig:gradual-abrupt-combined}(a), intermediate recalibration reduces the required robustness correction by approximately \(59\%\) compared with abrupt replacement, indicating a smaller effective distribution shift. In Fig.~\ref{fig:gradual-abrupt-combined}(b), both strategies exceed the nominal \(95\%\) coverage level, but gradual deployment achieves this with approximately \(36\%\) lower conservativeness. Overall, the results show that replacing control policies incrementally can preserve coverage while avoiding the over-conservatism caused by protecting against an abrupt one-step control policy change.

\providecommand{\datapath}{figures/gradual_vs_abrupt/}

\definecolor{rwthblue100}{HTML}{00549F}
\definecolor{rwthblue50}{HTML}{8EBAE5}
\definecolor{rwthpetrol100}{HTML}{0F4D58}
\definecolor{rwthgray75}{HTML}{7F7F7F}
\definecolor{rwthred100}{HTML}{CC071E}

\pgfplotstableread[col sep=comma]{\datapath/6_score_overlap_refs.csv}\scorerefs
\pgfplotstablegetelem{0}{q_cp}\of{\scorerefs}\edef\abruptqcp{\pgfplotsretval}
\pgfplotstablegetelem{0}{q_star}\of{\scorerefs}\edef\abruptqstar{\pgfplotsretval}
\pgfplotstablegetelem{0}{target_miss_at_q_cp}\of{\scorerefs}
\edef\abruptmiss{\pgfplotsretval}
\pgfplotstablegetelem{0}{x_min}\of{\scorerefs}\edef\scorexmin{\pgfplotsretval}
\pgfplotstablegetelem{0}{x_max}\of{\scorerefs}\edef\scorexmax{\pgfplotsretval}
\pgfplotstablegetelem{1}{q_cp}\of{\scorerefs}\edef\gradualqcp{\pgfplotsretval}
\pgfplotstablegetelem{1}{q_star}\of{\scorerefs}\edef\gradualqstar{\pgfplotsretval}
\pgfplotstablegetelem{1}{target_miss_at_q_cp}\of{\scorerefs}
\edef\gradualmiss{\pgfplotsretval}
\pgfmathsetmacro{\scoreplotxmax}{1.03*max(\scorexmax,max(\abruptqstar,\gradualqstar))}
\pgfmathsetmacro{\abruptmisspct}{100*\abruptmiss}
\pgfmathsetmacro{\gradualmisspct}{100*\gradualmiss}

\begin{figure}[t]
\centering
\begin{tikzpicture}
\begin{groupplot}[
    group style={group size=1 by 2, vertical sep=1.75cm},
    width=0.89\columnwidth,
    height=0.39\columnwidth,
    ymin=0,
    xmin=\scorexmin,
    xmax=\scoreplotxmax,
    xtick={0,2,4,6,8,10},
    minor x tick num=0,
    ylabel={Density},
    tick label style={font=\scriptsize},
    label style={font=\footnotesize},
    axis line style={black!65},
    tick style={black!65},
    legend style={
        font=\tiny,
        at={(0.98,0.97)},
        anchor=north east,
        draw=none,
        fill=white,
        fill opacity=0.88,
        text opacity=1,
        row sep=-1pt,
    },
]
\nextgroupplot[
     xlabel={Nonconformity score (p.u.)},
]
\addplot+[ybar interval, area legend, mark=none, fill=rwthgray75, draw=rwthgray75, opacity=0.50]
table[
    x=bin_left,
    y=density,
    col sep=comma,
    restrict expr to domain={\thisrow{series_code}}{0:0},
]{\datapath/6_score_overlap_hist.csv};
\addlegendentry{Source (calib.)}

\addplot+[ybar interval, area legend, mark=none, fill=rwthred100, draw=rwthred100, opacity=0.50]
table[
    x=bin_left,
    y=density,
    col sep=comma,
    restrict expr to domain={\thisrow{series_code}}{1:1},
]{\datapath/6_score_overlap_hist.csv};
\addlegendentry{3RL target}

\coordinate (abrupt-qcp-x) at (axis cs:\abruptqcp,0);
\coordinate (abrupt-qstar-x) at (axis cs:\abruptqstar,0);
\coordinate (abrupt-bottom) at (rel axis cs:0,0);
\coordinate (abrupt-top) at (rel axis cs:0,1);
\draw[black, dotted, thick]
    (abrupt-qcp-x |- abrupt-bottom) -- (abrupt-qcp-x |- abrupt-top);
\draw[black, thick]
    (abrupt-qstar-x |- abrupt-bottom) -- (abrupt-qstar-x |- abrupt-top);

\nextgroupplot[
    xlabel={Nonconformity score (p.u.)},
]

\addplot+[ybar interval, area legend, mark=none, fill=rwthgray75, draw=rwthgray75, opacity=0.50]
table[
    x=bin_left,
    y=density,
    col sep=comma,
    restrict expr to domain={\thisrow{series_code}}{10:10},
]{\datapath/6_score_overlap_hist.csv};
\addlegendentry{Source (calib.)}

\addplot+[ybar interval, area legend, mark=none, fill=rwthpetrol100, draw=rwthpetrol100, opacity=0.50]
table[
    x=bin_left,
    y=density,
    col sep=comma,
    restrict expr to domain={\thisrow{series_code}}{11:11},
]{\datapath/6_score_overlap_hist.csv};
\addlegendentry{3RL target}

\coordinate (gradual-qcp-x) at (axis cs:\gradualqcp,0);
\coordinate (gradual-qstar-x) at (axis cs:\gradualqstar,0);
\coordinate (gradual-bottom) at (rel axis cs:0,0);
\coordinate (gradual-top) at (rel axis cs:0,1);
\draw[black, dotted, thick]
    (gradual-qcp-x |- gradual-bottom) -- (gradual-qcp-x |- gradual-top);
\draw[black, thick]
    (gradual-qstar-x |- gradual-bottom) -- (gradual-qstar-x |- gradual-top);

\end{groupplot}

\node[font=\footnotesize, anchor=north]
    at ($(group c1r1.south)+(0,-0.95cm)$)
    {(a) Abrupt shift: droop to 3RL};

\node[font=\footnotesize, anchor=north]
    at ($(group c1r2.south)+(0,-0.95cm)$)
    {(b) Gradual shift: 2RL to 3RL};
\end{tikzpicture}
\caption{Source-calibration and 3RL-target score distributions. Dotted and solid lines mark $\tau_{\mathrm{CP}}$ and $\tau^*$, respectively. Target miss rates are \pgfmathprintnumber[fixed,precision=2]{\abruptmisspct}\% (abrupt) and \pgfmathprintnumber[fixed,precision=2]{\gradualmisspct}\% (gradual).}
\label{fig:score-overlap}
\end{figure}
Fig.~\ref{fig:score-overlap} compares the calibration and
3RL-target score distributions. The abrupt transition produces a more
pronounced discrepancy in the upper tail: \(7.36\%\) of the abrupt-transition test scores exceed the standard
conformal threshold, compared with \(5.62\%\) under gradual replacement. The latter is much closer to the nominal $5\%$ miscoverage
level. Overall, the results show that sequential data collection and
recalibration reduce the effective shift, resulting in a smaller robust
correction and tighter prediction intervals while maintaining empirical coverage above the nominal target

\begin{table}[t]
\centering
\caption{Warning performance under gradual control policy shift. Point-level metrics are computed over individual time points within the prediction horizon. Trajectory-level metrics indicate whether any time point within the horizon is violating or uncertain.}
\label{tab:warning_performance_controller_change}
\begin{tabular}{lcc}
\toprule
Metric & Point level & Trajectory level \\
\midrule
Violation rate & 0.0156 & 0.4200 \\
Alarm rate & 0.0965 & 0.7710 \\
Recall & 0.9965 & 1.0000 \\
Precision of alarms & 0.1610 & 0.5447 \\
\bottomrule
\end{tabular}
\end{table}
Table~\ref{tab:warning_performance_controller_change} reports the warning performance under gradual control policy shift. A point-level alarm occurs when the prediction interval at a given bus and time intersects or exceeds a voltage limit. A trajectory-level alarm occurs when at least one point-level alarm is raised within the horizon. Recall is the fraction of true violations that are alarmed, while precision is the fraction of alarms that correspond to true violations. Voltage violations are rare at the point level, with a violation rate of \(0.0156\), but the method detects almost all of them, achieving recall \(0.9965\). This comes with conservative alarms: \(9.65\%\) of point-level checks are flagged, and the corresponding precision is \(0.1610\). At the trajectory level, the method flags \(77.10\%\) of rollouts and achieves recall \(1.0000\), meaning that every rollout with at least one violation is detected. The trajectory-level precision is \(0.5447\), so more than half of the alarmed rollouts contain a true violation. Overall, the screening prioritizes avoiding missed voltage violations under control policy shift, at the cost of some conservative alarms.

\section{Conclusion}
\label{sec:conclusion}

This paper addresses a critical question: can we develop an efficient safety screening framework for new voltage control policies under imperfect simulations and historical data collected under another control policy? We solve this by proposing DR-CSS, which integrates nominal simulations with distributionally robust conformal screening to systematically bound both simulation-to-reality and policy-induced distribution shifts. Validated across multiple operational scenarios and distribution grid topologies, this framework demonstrates great potential, consistently maintaining tight, tight safety bounds where traditional importance-weighted baselines result in unbounded intervals. Ultimately, by uniting this rigorous methodological backbone with practical deployment policies, specifically a staged gradual replacement strategy, DR-CSS guarantees that safety screening of advanced data-driven control remains efficient and informative.

The results suggest several promising directions for future refinement, spanning both methodological enhancements and broader power system applications. One possible direction is to advance the underlying framework to continuous conformal risk control, shifting from a binary accept-or-reject threshold to the mathematical optimization and minimization of the expected physical severity or duration of minor overvoltages. In terms of application, DR-CSS can be expanded to evaluate multi-layered grid architectures where fast device controls and slow market scheduling interact across different time scales, as well as partially observable grids where operators must rely on imperfect state estimations. 
\emph{During the preparation of this work, the authors used ChatGPT and Claude to 
improve readability and assist with grammar and phrasing. After using this 
tool, the authors reviewed and edited the content as needed and take full 
responsibility for the content of the publication.}


\bibliographystyle{IEEEtran}
\bibliography{ref}

@article{ZHA21,
  author  = {Zhang, Ying and Wang, Xinan and Wang, Jianhui and Zhang, Yingchen},
  title   = {Deep Reinforcement Learning Based Volt-VAR Optimization in Smart Distribution Systems},
  journal = {IEEE Trans. Smart Grid},
  volume  = {12},
  number  = {1},
  pages   = {361--371},
  year    = {2021},
  doi     = {10.1109/TSG.2020.3010130}
}

@article{ANG22,
  author  = {Angelopoulos, Anastasios N. and Bates, Stephen},
  title   = {A Gentle Introduction to Conformal Prediction and Distribution-Free Uncertainty Quantification},
  journal = {arXiv preprint arXiv:2107.07511},
  year    = {2022},
  doi     = {10.48550/arXiv.2107.07511}
}

@article{TIB20,
  author  = {Tibshirani, Ryan J. and Barber, Rina Foygel and Candes, Emmanuel J. and Ramdas, Aaditya},
  title   = {Conformal Prediction Under Covariate Shift},
  journal = {arXiv preprint arXiv:1904.06019},
  year    = {2020},
  doi     = {10.48550/arXiv.1904.06019}
}

@article{TAU22,
  author  = {Taufiq, Muhammad Farhan and Ton, Jean-Francois and Cornish, Robert and Teh, Yee Whye and Doucet, Arnaud},
  title   = {Conformal Off-Policy Prediction in Contextual Bandits},
  journal = {arXiv preprint arXiv:2206.04405},
  year    = {2022},
  doi     = {10.48550/arXiv.2206.04405}
}

@inproceedings{KUI24,
  author    = {Kuipers, Tom and Tumu, Renukanandan and Yang, Shuo and Kazemi Mehrabadi, Milad and Mangharam, Rahul and Paoletti, Nicola},
  title     = {Conformal Off-Policy Prediction for Multi-Agent Systems},
  booktitle = {2024 IEEE 63rd Conference on Decision and Control (CDC)},
  pages     = {1067--1074},
  year      = {2024},
  doi       = {10.1109/CDC56724.2024.10886791}
}

@article{REN24,
  author  = {Renkema, Yvet and Brinkel, Nico and Alskaif, Tarek},
  title   = {Conformal Prediction for Stochastic Decision-Making of PV Power in Electricity Markets},
  journal = {Electric Power Systems Research},
  volume  = {234},
  pages   = {110750},
  year    = {2024},
  doi     = {10.1016/j.epsr.2024.110750}
}

@inproceedings{OTT18,
  author    = {Otte, M. and Leimgruber, F. and Brundlinger, R. and others},
  title     = {Hardware-in-the-Loop Co-Simulation Based Validation of Power System Control Applications},
  booktitle = {2018 IEEE 27th International Symposium on Industrial Electronics (ISIE)},
  pages     = {1229--1234},
  year      = {2018},
  doi       = {10.1109/ISIE.2018.8433761}
}

@article{PAS24,
  author  = {Paspatis, A. and Kontou, A. and Kotsampopoulos, P. and others},
  title   = {Advanced Hardware-in-the-Loop Testing Chain for Investigating Interactions Between Smart Grid Components During Transients},
  journal = {Electric Power Systems Research},
  volume  = {228},
  pages   = {109990},
  year    = {2024},
  doi     = {10.1016/j.epsr.2023.109990}
}

@article{Wang2020SafeOffPolicy,
  author  = {Wei Wang and Nanpeng Yu and Yuanqi Gao and Jie Shi},
  title   = {Safe Off-Policy Deep Reinforcement Learning Algorithm for Volt-VAR Control in Power Distribution Systems},
  journal = {IEEE Transactions on Smart Grid},
  volume  = {11},
  number  = {4},
  pages   = {3008--3018},
  year    = {2020},
  doi     = {10.1109/TSG.2019.2962625}
}

@article{Gao2021ConsensusMARL,
  author  = {Yuanqi Gao and Wei Wang and Nanpeng Yu},
  title   = {Consensus Multi-Agent Reinforcement Learning for Volt-VAR Control in Power Distribution Networks},
  journal = {IEEE Transactions on Smart Grid},
  volume  = {12},
  number  = {4},
  pages   = {3594--3604},
  year    = {2021},
  doi     = {10.1109/TSG.2021.3058996}
}

@article{Kabir2023TwoTimescaleVVC,
  author  = {Farzana Kabir and Nanpeng Yu and Yuanqi Gao},
  title   = {Deep Reinforcement Learning-Based Two-Timescale Volt-VAR Control With Degradation-Aware Smart Inverters in Power Distribution Systems},
  journal = {Applied Energy},
  volume  = {335},
  pages   = {120629},
  year    = {2023},
  doi     = {10.1016/j.apenergy.2022.120629}
}

@article{Gao2022ModelAugmentedSafeRL,
  author  = {Yuanqi Gao and Nanpeng Yu},
  title   = {Model-Augmented Safe Reinforcement Learning for Volt-VAR Control in Power Distribution Networks},
  journal = {Applied Energy},
  volume  = {313},
  pages   = {118762},
  year    = {2022},
  doi     = {10.1016/j.apenergy.2022.118762}
}

@inproceedings{Everett2021NNVerification,
  author    = {Michael Everett},
  title     = {Neural Network Verification in Control},
  booktitle = {2021 60th IEEE Conference on Decision and Control (CDC)},
  pages     = {6326--6340},
  year      = {2021},
  doi       = {10.1109/CDC45484.2021.9683154}
}

@inproceedings{Kochdumper2023OpenClosedLoop,
  author    = {Niklas Kochdumper and Christian Schilling and Matthias Althoff and Stanley Bak},
  title     = {Open- and Closed-Loop Neural Network Verification Using Polynomial Zonotopes},
  booktitle = {NASA Formal Methods},
  series    = {Lecture Notes in Computer Science},
  volume    = {13903},
  pages     = {16--36},
  publisher = {Springer},
  year      = {2023},
  doi       = {10.1007/978-3-031-33170-1_2}
}

@article{Tran2022VerificationLECPS,
  author  = {Hoang-Dung Tran and Weiming Xiang and Taylor T. Johnson},
  title   = {Verification Approaches for Learning-Enabled Autonomous Cyber-Physical Systems},
  journal = {IEEE Design \& Test},
  volume  = {39},
  number  = {1},
  pages   = {24--34},
  year    = {2022}
}

@article{Dawson2023LearnedCertificates,
  author  = {Charles Dawson and Sicun Gao and Chuchu Fan},
  title   = {Safe Control With Learned Certificates: A Survey of Neural Lyapunov, Barrier, and Contraction Methods for Robotics and Control},
  journal = {IEEE Trans. Robot.},
  volume  = {39},
  number  = {3},
  pages   = {1749--1767},
  year    = {2023},
  doi     = {10.1109/TRO.2022.3232542}
}

@article{Geth2023DSSEPractical,
  author  = {Frederik Geth and Marta Vanin and Werner Van Westering and Terese Milford and Amritanshu Pandey},
  title   = {Making Distribution State Estimation Practical: Challenges and Opportunities},
  journal = {arXiv preprint arXiv:2311.07021},
  year    = {2023},
  doi     = {10.48550/arXiv.2311.07021}
}

@article{Parlaktuna2024TopologyError,
  author  = {Kemal Parlaktuna and Erk Dursun and Murat G{\"o}l},
  title   = {WLAV State Estimation Based Topology Error Detection and Identification in Distribution Networks With Limited Number of Measurements},
  journal = {Electric Power Systems Research},
  volume  = {232},
  pages   = {110406},
  year    = {2024},
  doi     = {10.1016/j.epsr.2024.110406}
}

@article{Wang2025UncertaintyReview,
  author  = {Z. Wang and S. Q. Bu and J. Wen and C. Huang},
  title   = {A Comprehensive Review on Uncertainty Modeling Methods in Modern Power Systems},
  journal = {International Journal of Electrical Power \& Energy Systems},
  volume  = {166},
  pages   = {110534},
  year    = {2025},
  doi     = {10.1016/j.ijepes.2025.110534}
}

@article{Wang2019PHIL,
  author  = {Wang, Yu and Syed, Mazheruddin H. and Guillo-Sansano, Efren and Xu, Yan and Burt, Graeme M.},
  title   = {Inverter-Based Voltage Control of Distribution Networks: A Three-Level Coordinated Method and Power Hardware-in-the-Loop Validation},
  journal = {IEEE Trans. Sustain. Energy},
  volume  = {11},
  number  = {4},
  pages   = {2380--2391},
  year    = {2020},
  doi     = {10.1109/TSTE.2019.2957010}
}

@article{Nguyen2022OnlineSafeDRL,
  author  = {Nguyen, Hoang Tien and Choi, Dae-Hyun},
  title   = {Three-Stage Inverter-Based Peak Shaving and Volt-VAR Control in Active Distribution Networks Using Online Safe Deep Reinforcement Learning},
  journal = {IEEE Trans. Smart Grid},
  volume  = {13},
  number  = {4},
  pages   = {3266--3277},
  year    = {2022},
  doi     = {10.1109/TSG.2022.3166192}
}

@article{Renkema2024PVForecast,
  author  = {Renkema, Yvet and Visser, Lennard and AlSkaif, Tarek},
  title   = {Enhancing the Reliability of Probabilistic PV Power Forecasts Using Conformal Prediction},
  journal = {Solar Energy Advances},
  volume  = {4},
  pages   = {100059},
  year    = {2024},
  doi     = {10.1016/j.seja.2024.100059}
}

@article{MURRAY2021111100,
title = {Voltage control in future electrical distribution networks},
journal = {Renewable and Sustainable Energy Reviews},
volume = {146},
pages = {111100},
year = {2021},
issn = {1364-0321},
doi = {https://doi.org/10.1016/j.rser.2021.111100},
author = {William Murray and Marco Adonis and Atanda Raji}
}

@article{CUI2022108609,
title = {Decentralized safe reinforcement learning for inverter-based voltage control},
journal = {Electric Power Systems Research},
volume = {211},
pages = {108609},
year = {2022},
issn = {0378-7796},
doi = {https://doi.org/10.1016/j.epsr.2022.108609},
url = {https://www.sciencedirect.com/science/article/pii/S037877962200685X},
author = {Wenqi Cui and Jiayi Li and Baosen Zhang}
}

@ARTICLE{Saverio2019,
  author={Bolognani, Saverio and Carli, Ruggero and Cavraro, Guido and Zampieri, Sandro},
  journal={ IEEE Control Netw. Syst.}, 
  title={On the Need for Communication for Voltage Regulation of Power Distribution Grids}, 
  year={2019},
  volume={6},
  number={3},
  pages={1111-1123},
  doi={10.1109/TCNS.2019.2921268}}

@article{MOLLAALI2025107809,
title = {Conformalized prediction of post-fault voltage trajectories using pre-trained and finetuned attention-driven neural operators},
journal = {Neural Networks},
volume = {192},
pages = {107809},
year = {2025},
issn = {0893-6080},
doi = {https://doi.org/10.1016/j.neunet.2025.107809},
url = {https://www.sciencedirect.com/science/article/pii/S0893608025006896},
author = {Amirhossein Mollaali and Gabriel Zufferey and Gonzalo Constante-Flores and Christian Moya and Can Li and Meng Yue and Guang Lin}
}

@article{pareek2025learningpowerflowconfidence,
      title={Learning Power Flow with Confidence: A Probabilistic Guarantee Framework for Voltage Risk}, 
      author={Parikshit Pareek and Sidhant Misra and Deepjyoti Deka},
      year={2025},
      eprint={2308.07867},
      archivePrefix={arXiv},
      primaryClass={eess.SY},
      url={https://arxiv.org/abs/2308.07867}, 
}

@inproceedings{WANG2021,
author = {Wang, Jianhong and Xu, Wangkun and Gu, Yunjie and Song, Wenbin and Green, Tim C.},
title = {Multi-agent reinforcement learning for active voltage control on power distribution networks},
year = {2021},
isbn = {9781713845393},
publisher = {Curran Associates Inc.},
address = {Red Hook, NY, USA},
articleno = {250},
numpages = {14},
series = {NIPS '21}
}


\end{document}